\documentstyle[12pt]{article}

\title{BUSHES OF VIBRATIONAL MODES FOR FERMI-PASTA-ULAM CHAINS}
\author{G.M.Chechin, N.V.Novikova and A.A.Abramenko\\ 
{\small \it Department of Physics, Rostov State University,
Zorge 5, Rostov-on-Don, 344090 Russia}}
\date{}
\begin{document}
\maketitle
\hrule 
{\small {\bf Abstract\rule{0pt}{20pt}}
Some exact solutions and multi-mode invariant submanifolds 
were found for the Fermi-Pasta-Ulam (FPU) $\beta$-model by 
Poggi and Ruffo in Phys.~D {\bf103} (1997) 251. In the present paper 
we demonstrate how results of such a type can be obtained for an
{\it arbitrary} N-particle chain with periodic boundary conditions 
with the aid of our group-theoretical approach [Phys.~D {\bf117} 
(1998) 43] based on the concept of bushes of normal modes in
mechanical systems with discrete symmetry. The integro-differential
equation describing the FPU-$\alpha$ dynamics in 
the modal space is derived. The loss of stability of the bushes of 
modes for the FPU-$\alpha$ model, in particular, for the 
limiting case $N \to \infty$ for the dynamical regime with 
displacement pattern having period twice the lattice spacing ($\pi $-mode) is studied. Our 
results for the FPU-$\alpha$ chain are compared with 
those by Poggi and Ruffo for the FPU-$\beta$ chain.

 
}


\section{Introduction}
Fermi, Pasta and Ulam (FPU) introduced in \cite{l1} a simple nonlinear 
model for studying the problem of equipartition of energy among different 
degrees of freedom (linear normal modes) of an $N$-particle mechanical 
system for sufficiently large $N$. This model 
represents a monatomic chain with interactions only between
neighboring atoms. It may be also considered as a chain of identical
masses connected with identical nonlinear springs.
 The force ${F(\Delta x)}$ produced by each spring
 can be expressed as a power series in
its deformation $\Delta x$:
\begin{equation} \label{eq0}
F(\Delta x) = -k\cdot\Delta x + \alpha\cdot(\Delta x)^2 + 
\beta\cdot(\Delta x)^3 + ...		
\end{equation}
The FPU-$\alpha$ chain corresponds to the case
 $\alpha \ne 0,\;\beta = 0$, while the FPU-$\beta$ chain corresponds
 to the case $\alpha = 0, \;\beta \ne 0$, and in the both 
models all terms of higher order than those written in (\ref{eq0})
 are neglected.

	The FPU model played an important role in the development of some new 
concepts in the nonlinear dynamics of classical systems and helped to 
reveal a number of new nonlinear phenomena (see, for example,
 \cite{l2,l3,l26}
 and references cited therein). Let us specifically refer to the approximately 
recurrent behavior in time of the energy distribution among 
several of the first modes \cite{l1}, to the introduction of 
the concept of solitons \cite{l4}, to some important results 
concerning deterministic chaos \cite{l5}, to breathing self-localized 
solitons \cite{l6}, etc.

	An interesting paper devoted to the FPU-$\beta $ chain dynamics was 
published recently by Poggi and Ruffo \cite{l3}. These authors revealed some 
exact solutions for the considered mechanical system and demonstrated 
``the existence of subsets of normal modes where energy remains trapped 
for suitable initial conditions", which they
 called ``subsets of I-type"\footnote{We use another
 term for such objects - bushes of normal modes [7 - 10].}.    
They also discussed the problem of stability of such solutions and 
appropriate multi-mode invariant submanifolds (with particular attention 
to the case of the thermodynamical limit ${N \to\infty}$).

	In connection with the above mentioned paper we want to emphasize 
the following essential points:

$ \bullet$  The analysis in \cite{l3} is based  on the specific 
character of 
interaction between particles of the FPU-$\beta $ chain and, therefore, such 
an analysis must be done once again for every other type of
monatomic chain.

$ \bullet$ Only one- and two-dimensional subsets of normal 
modes of ``type I" were found for FPU-$\beta $   in \cite{l3}.

	In the present paper we would like to demonstrate a simple
group-theoretical method, based on our previous
 papers [7--10],   for finding all subsets of modes 
of ``type I" for monatomic chains with periodical boundary 
conditions. This method, in its general form, can be used for any 
mechanical system with any space symmetry group \cite{l10, l11}.\footnote{
In particular, it can be used for multiatomic chains.}
 It is based on the concept of ``bushes of normal modes"
in  nonlinear physical systems with discrete symmetry,
 which we consider to be as fundamental as the concepts 
of solitons, dissipative structures, etc. in modern nonlinear
 science. Apparently,  bushes of modes 
play an important role in many physical phenomena of current interest 
\cite{l10}. For example,

$ \bullet$   peculiarities of dielectric and optical spectra brought about by 
the 
interactions between different modes in crystals;

$ \bullet$   critical dynamics, behavior of structural and thermodynamical 
parameters near phase transition points;

$ \bullet$   phenomena induced by interaction between soft and hard modes, in 
particular, peculiarities of the temperature dependence of the 
Debye\,-Waller factor near phase transitions,
 which plays an important role in the interpretation
 of x-ray and neutron structural experiments, etc.

The present paper is devoted to the {\it symmetry-determined bushes of vibrational modes} in the 
monatomic chains. 
Let us explain these basic notions on the qualitative level.
The more exact definitions we give in the next section.

{\it Modes}. We consider the symmetry group $G_0$ of a given mechanical system in equilibrium 
(``parent group") and its irreducible representations (irreps).
The basis vectors of these irreps, which describe the atomic displacements patterns, are called 
``symmetry-adapted coordinates".
In a particular case (the monatomic chain is precisely this case!), symmetry-adapted coordinates can 
be identical with {\it normal} coordinates.
We mean by the term ``mode" an arbitrary superposition (with time-dependent coefficients)
of the above basis vectors of a given multidimensional irrep.
All that was just said allows us to use for the {\it monatomic} chain the term ``modes"
(or ``vibrational modes") as a synonym of both the symmetry-adapted and normal 
modes\footnote{
In more complicated cases when a given irrep enters several times into the full vibrational 
representation of the considered mechanical system, we must differentiate between the terms ``symmetry-
adapted" and ``normal" coordinates (modes). }.

In the most part of the paper, we treat $G_0$ as a cyclic group of pure translations (only in Sec. 2.6, we 
consider $G_0$ as the dihedral group).
All irreps of this group are one-dimensional, and each vibrational mode is simply a product of some 
time-dependent coefficient with the appropriate normal coordinate which can be introduced in the 
harmonic approximation by the conventional method.

{\it Bushes of modes}. Normal modes are independent of each other in the harmonic approximation, 
but the excitation from the initially excited mode (we call it ``root" mode) can spread to a number of
modes with zero amplitudes at the initial instant (we call them ``secondary" modes), if anharmonic 
terms in the  Hamiltonian of the considered mechanical system are taken into account.

It is essential that, in general, {\it not all modes} turn out to be excited as a result of the excitation of 
the given root mode, but only a very certain their collection. This phenomenon is a consequence of the 
specific selection rules for the excitation transfer between normal modes of different symmetry[7].
The bush of normal modes is a superposition of {\it all} modes, associated with different irreps of the 
parent group $G_0$, which are involved in the vibrational process as a result of exciting a given root mode.

Every bush describes an {\it exact} dynamical regime whose number of degrees of freedom can be 
essentially less than that of the considered mechanical system.\footnote{
Only in the special case, a given bush can possess the trivial symmetry and represent the general 
dynamical regime whose dimension coincides with that of the considered mechanical system.}
Note that one-dimensional bushes represent the similar nonlinear normal modes introduced by 
Rosenberg [15].

As was already stated, excitation of a primary (``root") 
mode leads to the excitation of the bush as a single dynamical object.
Amplitudes of modes belonging to the bush change in time and, as a 
rule, their evolution is not trivial. As a consequence,
we can speak about 
a number of new types of excitations in systems with discrete 
symmetry.

The following propositions were justified in the previous papers [7-10]:

1. A certain subgroup $G$ of the symmetry group $G_0$ corresponds to a given bush, and this bush 
can be excited by imposing the appropriate initial conditions with the above symmetry group 
$G \subset  G_0$.

2. Each mode belonging to the bush possesses its own symmetry group which is greater than or equal 
to the group $G$ of the whole bush.

3. In spite of evolving mode amplitudes, the complete collection of modes in the given bush is 
preserved in time and, in this sense, the bush can be considered as a geometrical object.

4. The energy of the initial excitation is trapped in the bush, i.e. it cannot spread to the modes which 
do not belong to the bush, because of the symmetry restrictions.

5. As an indivisible nonlinear object, the bush exists because of force interactions between the modes 
contained in it.

6. Taking into account the concrete type of interactions between particles of the considered 
mechanical system can only reduce the dimension of the given bush.

7. The extension of the bush can be realized as a result of the loss of its stability which is 
accompanied by spontaneous breaking of the bush symmetry (dynamical analog of phase 
transition).

In the present paper, we discuss {\it symmetry-determined} bushes of vibrational modes, i.e. bushes 
whose existence is brought about by the symmetry-related causes only.
In other words, we do not take into account any information about the type of interactions between the 
particles of the mechanical system.
Nevertheless, it should be noted that taking into account such information can lead to some {\it 
additional} selection rules for excitation transfer between different modes and, therefore, to the 
reduction of the dimension of the symmetry-determined bushes (we discuss this problem in 
Sec.2.5).
 Moreover, every additionally considered symmetry can result in a similar effect (see Sec.2.6.).
Thus, one can be sure that the excitation of a given mode cannot spread out of the 
symmetry-determined bush having that mode as root, but there may exist additional causes actually
restricting the excitation to a mode subset of this symmetry-determined bush.

Let us stress that the symmetry-determined bushes which are found in the present paper 
are valid for any of monatomic chains and, {\it in some a sense},
 they can be applied to multiatomic chains as well (we 
explain this point in Sec.2.6).

	The notion of bushes of normal modes was introduced in \cite{l7}. In 
the last paper and in [8, 9] we investigated the main properties of bushes 
of modes and developed the general group-theoretical method for their 
finding. The detailed discussion of our approach and some important 
theorems are presented in \cite{l10}. Bushes of small dimensionality for 
many structures with point and space groups of crystallographic 
symmetry are found, classified and investigated in the above papers. 
Bushes of vibrational modes for the fullerene $C_{60}$ (``buckyball" structure)
are discussed in  \cite{l23,l24}.

	Since application of group-theoretical methods for finding bushes 
of modes for mechanical systems, described by space groups with 
multidimensional irreducible representations, is very difficult, we 
developed a set of appropriate computer programs for treating 
such problems (previously similar programs were used for 
studying structural phase transitions in crystals \cite{l12}).

	All symmetry-determined ``resonance subspaces'' (see \cite{l13,l14}) and 
similar nonlinear normal modes
were found in \cite{l11} for {\it all} mechanical systems with any of the 230 space 
groups. Setting up this problem and its solution became possible because 
of using the concept of ``irreducible'' bushes of modes and due to the 
employment of the remarkable computer program ISOTROPY, 
created by Stokes and Hatch. This program realizes a great number 
of group-theoretical methods used in the theory of crystals. Its modified 
version, including the possibility to treat bushes of modes, is now
available on the Internet as free software \cite{l17}.

	The first part of the present paper is devoted to the problem of 
existence of bushes of normal modes in nonlinear chains, and the 
second part is devoted to the discussion of their stability. Considering the 
existence of a bush of modes in nonlinear periodical chains with arbitrary 
interaction between their atoms (not necessarily of nearest-neighbor 
type), we use the general group-theoretical method described in \cite{l10}. 
Discussing this problem we keep in mind not only the 
goal of obtaining some new results, but also the purpose of the 
exposition of our 
approach for studying dynamics of nonlinear systems with discrete 
symmetry using one-dimensional chains as the simplest case of such 
symmetry. Finally, we obtain for the FPU-$\alpha $ model a 
new integro-differential equation in the modal space for
 the case of the continuum limit (${N \to\infty}$). 
This equation is valid for all wave lengths (not only for
 long wavelengths as in the KdV-equation 
obtained for the FPU-$\alpha $ model by Zabusky and Kruskal \cite{l4}).

	In the second part of the paper, we discuss the problem of bush 
stability using the FPU-$\alpha $ chain as a concrete model 
(in contrast to the FPU-$\beta $ chain considered in \cite{l3}).
 Different channels of the loss 
of bush stability for finite  $N$  are discussed.
 The appearance of ``satellites'' forbidden 
by symmetry are revealed for bushes of normal modes for sufficiently
large values of $N$. The simplest bush $B[2a]$, corresponding to
the displacement pattern with periodicity twice 
the  periodicity of the chain in the equilibrium state\footnote{
This bush consists of only one zone boundary mode or $\pi$-mode, in another 
terminology.
}, is 
studied for the limit ${N \to\infty}$.

\section{Bushes of normal modes for nonlinear chains}
We describe here the general method for finding bushes of 
vibrational modes \cite{l10}  using the one-dimensional FPU model 
with  periodical boundary conditions as an illustrative 
example. The starting point of our approach is the expression 
of the set of atomic displacements as a sum of contributions 
from different irreducible representations (irreps) of the 
symmetry group of the considered mechanical system in equilibrium [see 
(\ref{eq2}) below]. Therefore, we begin with the consideration of 
irreps and their basis vectors for the case of the monatomic 
chain.

\subsection{Vibrations of monatomic chain in terms of 
irreducible representations of its symmetry group}

Let us consider the many-particle Hamiltonian system moving near the 
single equilibrium state characterized by the parent group $G_0$ of 
point or space symmetry  and introduce the $N$-dimensional 
vector  ${\bf  X}(t)$ determining values of all of its $ N $ degrees
 of freedom at a given instant  $t$ :
\begin{equation} \label{eq1}
{\bf  X}(t)= \{x_1(t),x_2(t),\ldots,x_N(t)\}.
\end{equation}
We can write the vector ${\bf  X}(t)$  as the superposition of basis
 vectors $\mbox{\boldmath$\varphi$}_k^{(j)}$ of all those irreducible
 representations $ \Gamma _j$ of the 
group $G_0$ which enter into the mechanical (reducible) 
representation of the considered system:
\begin{equation} \label{eq2}
{\bf  X}(t)= \sum_{jk}\mu_k^{(j)}(t)\cdot \mbox{\boldmath$\varphi$}_k^{(j)}=
\sum_{j}{\bf \Delta}_j(t)\;. 
\end{equation}
Here the index  $j$  corresponds  to the irrep $\Gamma _j$ while the index $k$  
corresponds to its different basis vectors because, in the general 
case, $\Gamma _j$ can be a multidimensional representation. The $N$-dimensional
vector ${\bf \Delta}_j(t)$ is the contribution to ${\bf  X}(t)$ from the irrep 
$\Gamma_j$.

	For an $N$-particle nonlinear {\it chain}, ${\bf  X}(t)$ is the complete 
set of displacements $x_i(t)(i = 1,2,\ldots, N)$  of each particle (mass 
point) from its equilibrium position\footnote{We consider longitudinal 
vibrations of the chain.}.
 [In accordance with the periodical boundary condition, we assume 
$x_{N+1}(t)\equiv x_1(t)$].

The translational group $T$ can be chosen as a simplest variant of the parent 
symmetry group $G_0$. Nevertheless, the monatomic chain is invariant under
 the action of {\it inversion} with respect to its center and, therefore, 
the {\it full} symmetry in this case is
the dihedral group $D$. Certainly, this group is more relevant for
the symmetry-based analysis of nonlinear vibrations in monatomic chains. 
On the other hand, the inversion can be absent for some {\it multiatomic}
chains, and the full symmetry for such cases is represented by the group $T$. 
Since some important results can be obtained for any one-dimensional chain
 with the aid of the group $T$ only, we start our consideration with
the more general and simple case $G_0=T$. The influence of the different 
choice of the parent group ($G_0=T$ or $G_0=D$) on the 
group-theoretical analysis of nonlinear vibrations of {\it monatomic} 
chains will be considered in Sec. 2.6.

Thus, we suppose that the equilibrium state of our monatomic chain is
described by the parent group $G_0=T$ which,  obviously, happens
 to be the cyclic group of order $N$ with 
generator $\hat a$:
\begin{equation} \label{eq3}
G_0=\{E,\hat a, \hat a^{2},\hat a^{3},\ldots, \hat a^{N-1}\},\mbox{    }
\hat a^{N}=E.
\end{equation}
($E$ is the identity element of the group $G_0$). Here $\hat a$ is the 
operator  translating all particles by the space period $ a$   
of the considered chain in equilibrium. 
This operator generates the cyclic permutation of 
all particles and, therefore, the permutation of the corresponding
displacements $x_i(t)$:
\begin{equation} \label{eq4}
\hat a\{x_1(t),x_2(t),\ldots\,x_{N-1}(t),x_N(t)\}=
\{x_N(t),x_1(t),x_2(t),\ldots\,x_{N-1}(t)\}\,.
\end{equation}

	As is well known \cite{l18}, the cyclic group $G_0$ of order 
$N$ has  $N$ irreps and they are all one-dimensional\footnote
{In contrast to many point and space groups 
which can possess multidimensional irreps.}.
Moreover, the $1\times 1$ matrices  of its generator
 $\hat a$ for different irreps $\Gamma_k\; (k=1,2, ..., N)$
 are simply the $N$th-degree roots of
1. As an  example, we give all matrix irreps of the cyclic group 
$G_0$ for $N=12$ in Table 1\footnote {Note, that in this table
and in some points of the text we neglect the hat over symbol $a$
for simplification of the notation.}\label{fn}. (Since all $\Gamma _k$ 
are one-dimensional 
irreps, Table 1 coincides with the table of their characters).

	Let us construct the basis vectors of the irreps 
of the group $G_0$ in (4) which describe the different modes of 
vibration of our nonlinear chain. The basis vectors of the irreps
are usually obtained by the 
method of the projection operator \cite{l18}, but in our case it is 
simpler to make use of the ``direct'' method 
based on the definition of the group representation (see \cite{l10}). 
Indeed, according to this definition
\begin{equation} \label{eq5}
\hat a\mbox{\boldmath$\Phi$}= \tilde M(a){\bf\Phi}\,,
\end{equation}
where $M(a)$ is the matrix corresponding in the irrep $\Gamma _k$ 
to an element $\hat a$ of the group $G_0$, and ${\bf  \Phi} $ 
is the complete set of basis vectors $\mbox{\boldmath$\varphi$}_k$  of this 
representation which are transformed under the action of the 
operator $\hat a$ [in (\ref{eq5}) we use tilde as a 
symbol of matrix transposition].

The matrix $M_k(a)$ of the generator 
$\hat a$ of the cyclic group $G_0$ for the irrep 
$\Gamma _k\;(k=0,1,2,\ldots ,N-1)$ is equal to
\begin{equation} \label{eq15}
M_k(a)=\gamma ^k\,,
\end{equation}
where $\gamma =e^{2\pi i/N}$ and, therefore, $\gamma ^N=1$.

The desired basis vector $\mbox{\boldmath$\varphi$}_k$ of the
 irrep $\Gamma_k$ can be written in  the form
\begin{equation} \label{eq100n}
\mbox{\boldmath$\varphi$}_k = (x_1^{(k)}, x_2^{(k)},\ldots, x_N^{(k)}),
\end{equation}
where $x_i^{(k)}$ ($i=1,\ldots,N$) are unknown displacements of the particles.
 From (\ref {eq5}) and (\ref{eq15}), we get the following equation which can be used for obtaining
 the basis vector $\mbox{\boldmath$\varphi$}_k$ :
\begin{equation} \label{eq101n}
\hat a \mbox{\boldmath$\varphi$}_k = \gamma ^k  \mbox{\boldmath$\varphi$}_k ,
\;\;\;\;\; k=0,1,\ldots,N-1.
\end{equation}
Equation (\ref{eq101n}) leads us to a chain of simple algebraic 
equations for the components of the vector $\mbox{\boldmath$\varphi$}_k$ 
 in (\ref{eq100n}).
Solving these equations, we obtain
\begin{equation} \label{eq102n}
\mbox{\boldmath$\varphi$}_k = (x, \;\bar  \gamma  ^k\cdot x,\; \bar  \gamma  ^{2k}\cdot x, 
\:\bar  \gamma  ^{3k}\cdot x, \ldots ,\;\bar  \gamma  ^{2k}\cdot x),
\end{equation}
where $x$ is an arbitrary constant determining the displacement of the first 
particle of the chain. (Hereafter, we denote complex conjugation by a bar over the 
appropriate  value, for example, $\bar \gamma =e^{-2\pi i/ N}$). The presence of only 
one arbitrary constant in the general
 form of each basis vector $\mbox{\boldmath$\varphi$}_k $ means that 
every irrep $\Gamma_k$ of the group $G_0$
enters one and only one time into the decomposition of the mechanical 
representation of the monatomic chain.

	The above conclusion applies to the monatomic chain, but it 
is not valid for arbitrary physical systems: a given irrep may 
enter several times or not enter at all into the appropriate 
mechanical representation. The method of finding the basis 
vectors for multidimensional irreps and complex mechanical 
structures turns out to be very complicated and it is useful 
to use in this case the computer  programs \cite{l12,l17}.

Normalizing the vector $\mbox{\boldmath$\varphi$}_k $ from Eq.~(\ref{eq102n}),
 we obtain the final form of the basis vectors of the irreps
 $\Gamma_k$ of the group $G_0=T$:
\begin{equation} \label{eq16}
\mbox{\boldmath$\varphi$}_k\equiv \mbox{\boldmath$\varphi$}\;(\Gamma_k)=\frac{1}{\sqrt N}
(1,\bar \gamma ^k,\bar \gamma ^{2k},\bar \gamma ^{3k},
\ldots,\bar \gamma ^{(N-1)k})\,.
\end{equation}
These basis vectors provide us so-called {\it symmetry-adapted} coordinates.
According to the Wigner theorem, they are {\it normal coordinates} 
associated with $\Gamma_k$, if only one copy of this irrep (multidimensional,
in general case) is contained in the decomposition of the mechanical representation 
into its irreducible parts\footnote{ Let us remember that in the opposite case, 
according to the Wigner theorem, additional linear transformation is required to 
diagonalize the force constant matrix. 
In our general approach we use symmetry-adapted coordinates since
they can be obtained with the aid of the group-theoretical methods alone
(without any information about concrete interactions in the mechanical 
system).}.

As was shown above, the two notions $-$ symmetry-adapted coordinates and normal coordinates
$-$  are identical when we characterize the monatomic chain by the translation group 
($G_0=T$).
We will refer to the modes determined by the above coordinates as normal 
modes or vibrational modes\footnote{ Modes differ from generalized coordinates
 $\mbox{\boldmath$\varphi$}_k $ 
by the appropriate time-dependent functions $\mu _k(t)$ in front of the vectors
 $\mbox{\boldmath$\varphi$}_k $ as in Eq.~(\ref{eq20}). 
In further consideration, we will sometimes call by the 
term ``modes" not only $\mu _k(t)\cdot \mbox{\boldmath$\varphi$}_k $, 
but also the vectors $\mbox{\boldmath$\varphi$}_k $ and even the functions $\mu _k(t)$.}.

Each basis vector $\mbox{\boldmath$\varphi$}_k $ determines a certain
 pattern of displacements of all $N$ atoms of the chain and, in some cases,
 it is more convenient to consider real modes $\mbox{\boldmath$\psi $}_k $
instead of the complex modes $\mbox{\boldmath$\varphi$}_k $.

Note that two basis vectors,  for which the sum of their indices is 
equal to $N$, are complex conjugates of each other:
\begin{equation} \label{eq18}
\mbox{\boldmath$\varphi$}_{N-k}=\overline{\mbox{\boldmath$\varphi$}_k},\;\;\;
 k=0,1,2,\ldots,\left[\frac{N}{2}\right]\;.
\end{equation}
(Hereafter, $\left[\frac{N}{2}\right]$ denotes the integer part of 
$\frac{N}{2}$.) Indeed,
$$
\mbox{\boldmath$\varphi$}_{N-k}=\frac{1}{\sqrt{N}}(1,\bar \gamma ^{N-k},
\bar \gamma ^{2(N-k)},\ldots)=\frac{1}{\sqrt{N}}(1,\bar \gamma ^{-k},
\bar \gamma ^{-2k},\ldots)=\frac{1}{\sqrt{N}}(1,\gamma ^k,
\gamma ^{2k},\ldots)=\overline {\mbox{\boldmath$\varphi$}_k}\,,
$$
since $\bar \gamma ^N=1$ and $\bar \gamma ^{-k}=\gamma ^k$.

Using the appropriate linear transformation in the space of 
these mutually conjugated vectors,  we obtain the real normal modes in the
 form given by Poggi and Ruffo \cite{l3}:
\begin{equation} \label{eq19}
\mbox{\boldmath$\psi $}_k = \left.\left\{\frac{1}{\sqrt{N}}\left[\sin\left(
\frac{2\pi k}{N}n \right)+ \cos\left(\frac{2\pi k}{N}n \right)
\right]\right|\;\;\;n=1,2,\ldots,N\,\right\}.
\end{equation}
Here $n$ is the number of the particle, and $k\;(k=0,1,2,\ldots,N-1)$  
is the number of the mode. We call the real vectors $\mbox{\boldmath$\psi $}_k$ 
and 
$\mbox{\boldmath$\psi $}_{N-k}$ ``conjugate modes'' similar to
the term for pairs of complex modes $\mbox{\boldmath$\varphi$}_k$ and
$\mbox{\boldmath$\varphi$}_{N-k}$.

	Returning to the basic Eq.~(\ref{eq2}), we can rewrite it in the 
following form, 
\begin{equation} \label{eq20}
{\bf X}(t)=\sum_{k=0}^{N-1}\mu _k(t)\mbox{\boldmath$\varphi$}_k=
\sum_{k=0}^{N-1}\nu _k(t)\mbox{\boldmath$\psi $}_k\,.
\end{equation}
Thus, the pattern of atomic displacements ${\bf X}(t)$ for the chain
in the vibrational regime at any instant $t$ can be represented as a 
superposition of complex ($\mbox{\boldmath$\varphi$}_k$) or real 
($\mbox{\boldmath$\psi $}_k$) basis vectors with the coefficients $\mu _k(t)$ or 
$\nu _k(t)$ which depend on the time $t$.
 
	Let us emphasize  that $\mbox{\boldmath$\psi $}_k$ from
 Eq.~(\ref{eq19}) are normal modes of the {\it linear}
 chain (they can be obtained in 
the harmonic approximation), but we use them in (\ref{eq20}) as the 
basis of the configuration space for studying dynamics of the 
{\it nonlinear} chain.

\subsection {Bushes of normal modes}

	We continue to adapt the general method  [7--11] of 
studying nonlinear dynamics of mechanical systems with 
arbitrary discrete symmetry for the case of nonlinear 
chains.
	Every basis vector (normal mode) from Eqs.~(\ref{eq16}) or
 (\ref{eq19}) determines a specific set
 of atomic displacements, and this pattern is characterized
 by a certain symmetry group $G_k$ which is a subgroup of the 
group $G_0$ of the chain in equilibrium ($G_k \subset G_0$).
 Let us excite a dynamical regime of the considered system using the set 
of atomic displacements corresponding to a given root mode 
$\mbox{\boldmath$\psi $}_k$ as initial conditions at the instant $t=0$:
\begin{equation} \label{eq21}
x_i\!\mid\!_{t=0}=A\; \psi _{ki}\,;\;\;\;\dot x_i\!\mid\!_{t=0}
=0\;\;\;(i=1,2,\ldots,N)\,.
\end{equation}
Here $\psi _{ki}$ are components of the vector $\mbox{\boldmath$\psi $}_k=
(\psi_{k1}, \psi_{k2},\ldots, \psi_{kN})$
and $A$ is a fixed constant.

 The symmetry group of the mechanical system at an 
arbitrary later moment $t$ cannot be lower than the group 
$G_k$ at the initial moment determined by the
vector $\mbox{\boldmath$\psi $}_k$.
Indeed, the symmetry of the system at some instant $t>t_0$ is determined by the
intersection of the symmetry  of its Hamiltonian and the symmetry of the initial 
conditions. On the other hand, $G_0$ can be considered as a symmetry group of 
the Hamiltonian, as well as of the equilibrium state, if this state is 
nondegenerate\footnote{ We state this hypothesis at the beginning of Sec. 2.1.
For the degenerate equilibrium state, i.e. when there exist
several equivalent minimums of the potential energy, the symmetry group
of the Hamiltonian is higher than that of the each of the appropriate equilibrium states.}.
Thus, every symmetry element $g \in  G_k$ does not change the dynamical 
state ${\bf X}(t)$  for $t > t_0$ (a more detailed consideration 
of this problem can be found in [7,8]).

Now we can pose the question: what modes $\mbox{\boldmath$\psi $}_k$
  from Eq.~(\ref{eq20}) can contribute to the 
considered dynamical regime ${\bf X}(t)$ corresponding to the group $G_k \subset  G$ for 
$t > t_0$?
 Obviously, because of nonlinear interactions between
 different modes, the root mode $\mbox{\boldmath$\psi $}_k$      
brings about the appearance of a number of ``secondary'' 
modes in the considered dynamical regime, and we can obtain all these modes by 
demanding that the  symmetry ${\bf X}(t)$ is equal to $G_k $ (at least, it  must 
not be lower than 
$G_k$!):
\begin{equation} \label{eq22}
\hat G_k\,{\bf X}(t)={\bf X}(t)\,.
\end{equation}
This equation means that ${\bf X}(t)$ is an {\it invariant}
vector with respect to all elements of the group $G_k$
\footnote{ Rigorously speaking, we must distinguish between the symmetry elements
 $g\subset G_k$, which act on three-dimensional vectors of the Euclidean space, 
and the {\it operators} $\hat g$, corresponding to them, which act 
on $N$-dimensional vectors ${\bf X}(t)$ (see [10]). 
The complete set of the operators $\hat g$ associated with 
all $g \in G_k$ forms the group $\hat G_k$.}.
As a consequence, only those modes $\mbox{\boldmath$\psi $}_j$ contribute to
${\bf X}(t)$ whose symmetry group $G_j$ is {\it higher than or equal to} 
$G_k$ of the initial configuration of the mechanical 
system in hand. The complete set of these modes, i.e., the root 
mode and all secondary modes corresponding to it, forms the 
{\it bush as the geometrical object}. 

Note that  {\it each} subgroup $G'\subset G_0$ must be tried 
in finding bushes, but in general, some subgroups do not generate any 
{\it vibrational} bushes (see examples in Sec. 2.6).

Let us consider the symmetry group $G_k$ of different modes
$\mbox{\boldmath$\psi $}_k$ in more detail. We can find the group 
$G_k\subset G_0$ of a given mode $\mbox{\boldmath$\psi $}_k$       
by selecting all elements $g\in G_0$ under which this mode is invariant:
\begin{equation} \label{eq23}
\hat g\,\mbox{\boldmath$\psi $}_k=\mbox{\boldmath$\psi $}_k\,.
\end{equation}
But there exists a more convenient method. Indeed, ${\bf X}(t)$      
was written in (\ref{eq2}) as a sum of contributions ${\bf\Delta }_j$
 from different irreps $\Gamma _j$ of the dimension $n_j$. 
It is easy to show that 
the following consequence of Eq.~(\ref{eq22}) takes place for each 
individual irrep $\Gamma_j$:
\begin{equation} \label{eq24}
\hat G_k\,{\bf \Delta }_j = {\bf \Delta }_j\,.
\end{equation}

According to the definition (\ref{eq5}) of the representation of the group 
$G_0$ we can act on the set of basis vectors
$\mbox{\boldmath$\psi $}_k^{(j)}$ by the matrix $\tilde M(g)$,
 instead of acting
on them by the operator $\hat g$ ($g\in G_0$). 
This is a trivial procedure for the cyclic group $G_0$, since all of
its irreps are one-dimensional and, therefore, these matrices are 
scalar values.\footnote{In the general case, the appropriate
 procedure can be 
realized using the notion of ``invariant'' vectors of 
the irreps of the group $G_0$.} 

On the other hand, Eq.~(\ref{eq23}) can be true if and only 
if the $1\times 1$ matrix $M(g)$, corresponding to the element
$g\in G_0$, is equal to the unit matrix. 
Therefore, all elements $g\in G_0$ whose matrices of
the irrep $\Gamma_k$ satisfy the condition
\begin{equation} \label{eq25}
M_k(g)=1
\end{equation}
do not change the mode $\mbox{\boldmath$\psi $}_k$ and, obviously,  the complete 
set 
of such elements forms the subgroup $G_k$ of the group 
$G_k\subset G_0$.

Let us consider the case $N=12$ as an illustrative example. All 
irreps of the corresponding cyclic group $G_0$ are given in 
Table 1, as well as the symmetry groups $G_k$ of modes 
$\mbox{\boldmath$\varphi$}_k$ 
associated with these irreps. As was explained above, only those 
elements $g\in G_0$ belong to the group $G_k$ whose matrices in 
the irrep $\Gamma_k$ are equal to 1.
In Table 1, we define the subgroups $G_k\subset G_0$ by putting their
  generators in square brackets. Hereafter, we write the 
element $a^p$ as $(p\cdot a)$: such notation is 
often used for cyclic groups. Note, that according to the 
well-known Lagrange theorem, the order ($M$) of any 
subgroup is a {\it divisor} of the order ($N$)  of this group, 
and that every subgroup of the cyclic group is also a cyclic 
group which can be determined by the single generator. Irreps $\Gamma _{N-k}$ 
and $\Gamma _k$ are mutual conjugate representations: $\Gamma _{N-
k}=\overline\Gamma _k$.  It is easy to notice that {\it all} subgroups $G_k$ of 
the original cyclic group $G_0$ with the order $N=12$ are presented
in Table 1, and that the subgroups corresponding to the conjugate irreps 
$\Gamma_k$ and $\Gamma_{N-k}$ are identical (more exactly, subgroups 
corresponding to the conjugate modes $\mbox{\boldmath$\varphi$}_k$ and 
$\mbox{\boldmath$\varphi$}_{N-k}$ associated with these irreps). 
Below we list subgroups $G_k$ of the group 
$G_0$ with $N=12$,  listing in braces all their elements (see
 footnote 8 on page \pageref{fn}),
the order $(M)$ and  symbols of irreps whose basis vectors 
(modes)  possess such a symmetry $(G_k)$.
\begin{enumerate}
\item $G_0[a]=\{E,a,2a,3a,\ldots,11a\},\;\;M_0=N=12$.

Only one mode $\mbox{\boldmath$\varphi$}_0 = 
\frac{1}{\sqrt{12}}(1,1,1,\ldots,1)$, corresponding to the irrep 
$\Gamma_0$, possesses this symmetry group. The mode 
$\mbox{\boldmath$\varphi$}_0$ plays a special role: it corresponds to the 
movement of the chain as  a whole. In the next sections,  this mode is not taken
into account since we consider only {\it vibrational} modes.

\item $G_1[2a]=\{E,2a,4a,6a,8a,10a\},\;\; M_1=6$.

It can be found from Table 1 that only two modes 
$\mbox{\boldmath$\varphi$} _0$ and $\mbox{\boldmath$\varphi$}_6$ 
satisfy such symmetry, because the unit matrix (1) 
corresponds to the generator [2a] {\it only} for the irreps 
$\Gamma_0$ and $\Gamma_6$. We write this fact as 
follows $$G_1[2a]:\Gamma_0, \Gamma_6 \,.$$
The symmetry group of the basis vector $\mbox{\boldmath$\varphi$}_6$
 is equal to $G_1[2a]$, while that of the basis vector
 $\mbox{\boldmath$\varphi$}_0$ is higher (it is equal to $G_0[a]$).
 Therefore, $\mbox{\boldmath$\varphi$}_6$ is the root mode,
and $\mbox{\boldmath$\varphi$}_0$ is a secondary mode corresponding 
to it.

	Proceeding  in a similar way, we obtain:

\item $G_2[3a]=\{E,3a,6a,9a\},\;\; M_2=4 : \;\;
\Gamma_0, \Gamma_4, \Gamma_8$.

\item $G_3[4a]=\{E,4a,8a\},\;\;M_3=3 : \;\;
\Gamma_0, \Gamma_3, \Gamma_6, \Gamma_9 $.

\item $G_4[6a]=\{E,6a\},\;\;M_4=2 : \;\;
\Gamma_0, \Gamma_2, \Gamma_4, \Gamma_6,\Gamma _8,\Gamma _{10} $.

\item$G_5[12a]=\{E\},\;\;M_5=1 : \;\;
\Gamma_j\;\;$ for $\;\;j=0,1,2,\ldots,11$.
\end{enumerate}
	Now  we can write all possible bushes for the chain with 
$N=12$  particles using Eq.~(\ref{eq20}) and the above results. 
Each bush corresponds to the appropriate subgroup  $G_k$ 
of the parent group $G_0$, and it embraces all modes with 
symmetry equal to (for root modes)  or higher than (for 
secondary modes) $G_k$.
\begin{eqnarray} \label{eq26}
G_1[2a]:\;\; {\bf x}^{(1)}(t)&=&\mu _0(t)\mbox{\boldmath$\varphi$}_0+
\mu _6(t)\mbox{\boldmath$\varphi$}_6,\nonumber\\
G_2[3a]:\;\;{\bf x}^{(2)}(t) &=&\mu _0(t)\mbox{\boldmath$\varphi$}_0+
\mu _4(t)\mbox{\boldmath$\varphi$}_4+\mu _8(t)\mbox{\boldmath$\varphi$}_8,\\
G_3[4a]:\;\; {\bf x}^{(3)}(t)&=&\mu _0(t)\mbox{\boldmath$\varphi$}_0+
\mu _3(t)\mbox{\boldmath$\varphi$}_3+\mu _6(t)\mbox{\boldmath$\varphi$}_6+
\mu _9(t)\mbox{\boldmath$\varphi$}_9,\nonumber\\
G_4[6a]:\;\; {\bf x}^{(4)}(t)&=&\mu _0(t)\mbox{\boldmath$\varphi$}_0+
\mu _2(t)\mbox{\boldmath$\varphi$}_2+\mu _4(t)\mbox{\boldmath$\varphi$}_4+
\mu _6(t)\mbox{\boldmath$\varphi$}_6+\mu _8(t)\mbox{\boldmath$\varphi$}_8+
\mu _{10}(t)\mbox{\boldmath$\varphi$}_{10},\nonumber\\
G_5[12a=E]:\;\;{\bf x}^{(5)}(t)&=&\sum_{j=0}^{11}
\mu _j(t)\mbox{\boldmath$\varphi$}_j, \nonumber \;\;
\mbox{(the trivial bush)}\,.
\end{eqnarray}

Thus there exist four {\it nontrivial} vibrational bushes (besides 
the trivial bush with $G_5[12a]=E$) for the nonlinear 
chain with $N=12$  particles. The modes (basis vectors of 
appropriate irreps) $\mbox{\boldmath$\varphi$}_k$ determine the pattern
of  atomic displacements, and the time-dependent 
coefficients $\mu_k(t)$ 
\footnote{Note, that $\mu_k(t)$            
 associated with different bushes are not related.
Indeed, we may introduce the additional index  $i$ for 
these coefficients $\mu_k^{(i)}(t)$, which corresponds to the number of 
the bush, but this index is omitted for simplification of the 
notation.} 
characterize the evolution of these displacements in time.

	It easy to find from (\ref{eq26}) that every bush with symmetry 
group $G_k$ contains the contributions from 
modes with this symmetry $G_k$ and from all other modes whose 
symmetry is {\it higher} than $G_k$. This is the {\it 
geometrical} cause of the spreading of the initial excitation 
from the root mode to other (secondary) modes with 
$G_k$ equal to or higher than that of the root mode. The {\it 
dynamical} reason for such a phenomenon will be considered 
below in Sec. 2.3.

	Returning to the general case, we obtain that the numbers 
and the types of possible bushes for any nonlinear chain 
with $N$ particles depend essentially on divisibility 
properties of the integer $N$. Indeed, the 
total number of bushes for an $N-$particle chain is equal to 
the number of divisors of $N$. The number of modes 
contained in the concrete bush corresponding to the divisor 
$M$ (this divisor determines the order of the appropriate 
cyclic group $G_k$) is equal to $N/M .$\footnote{Moreover, it
 is easy to verify that the order $M$ of the group $G_j[ma]$ 
equals $N/m$ and, therefore, the number of modes of the bush
 with such symmetry is simply equal to $m$.} In particular, for 
prime number $N$ there exists only one (trivial) bush with $G_1=\{E\}$
which contains all modes $	\mbox{\boldmath$\varphi$}_j$
 $(j=0,1,2,..., N-1)$.\footnote{ In this case of trivial symmetry, the bush
describes the full dynamics of the considered mechanical system
without any simplification.}
Therefore, the bush structure for the chain with $N=12$ differs in 
principle from that for $N=11$ or $N=13$.

In conclusion, let us note that all possible symmetry-determined bushes 
of normal modes can be found for arbitrary mechanical system with discrete symmetry
by the regular group-theoretical methods described in detail in our previous papers 
(see \cite{l10} and \cite{l11}). We will not consider these methods in the present
paper, but we want to extract their main points.

A certain subgroup $G$ of the parent group $G_0$ corresponds to a given
bush $B[G]$. This subgroup must be fixed for singling out the complete collection of the
normal modes contained in $B[G]$.\footnote{
In particular, the subgroup $G$ can be fixed by indication of the initially excited 
root mode, since the definite group is associated with each normal mode.}
Then the structure of the given bush, i.e. the list of all modes 
belonging to $B[G]$, can be found as a result of solving the following systems of linear 
algebraic equations for {\it each of the irreps} $\Gamma_j$ of the group $G_0$:

\begin{equation}\label{eq500}
(\Gamma _j \downarrow  G) \;{\bf C}_j = {\bf C}_j .
\end{equation}

Here $\Gamma_j \downarrow G$ is the restriction of the irrep $\Gamma_j$ of
the parent group $G_0$ to the subgroup $G$ of this group. In other words, we must solve
the equations  

$$
M_j (g_i) \;  {\bf C}_j = {\bf C}_j 
$$
for all matrices $M_j ( g_i )$ corresponding to the generators $g_i$ of the
subgroup $G$ in the irrep $\Gamma_j$.
The {\it invariant vectors} ${\bf C}_j $ from (\ref{eq500}) determine the set of coefficients in
the superposition of the basis vectors of the given irrep $\Gamma_j$ which is 
compatible with the symmetry group $G$ of the given bush $B[G]$. 
In particular, the invariant vectors can turn out to be zero vectors
for some irreps $\Gamma_j$, and this means that such irreps
do not contribute to the given bush.

For obtaining the bushes of {\it vibrational} modes\footnote{
 There exist bushes of modes of very different physical nature.}, 
in addition to the above results, we must find the explicit form of the basis vectors 
constructed from the set of atomic displacements, for all irreps which 
contribute to the given bush. (This can be done
for example, with the aid of the method of projection operators).

As we already noted, the above procedures are simple for the monatomic chain, 
but special group-theoretical computer programs are needed for the general case 
(see references \cite{l12} and \cite{l17}).

If the complete list of modes of a given bush is known, the {\it root} mode can be 
frequently defined as that with the lowest symmetry of all other bush modes. 
Nevertheless, there exist cases where the root mode of the bush can be chosen
in different ways and there exist cases where the symmetry group of the bush is 
determined by the {\it intersection} of the symmetry groups of the modes associated
with {\it several} different irreps of the parent group. In the last case, we speak about 
a bush with several root modes (see \cite{l10} and, especially, \cite{l24} and \cite{l25}).
In general, the set of root modes must uniquely determine the symmetry group
of the whole bush. The excitation of this set of root modes leads by 
necessity to the excitation of all other modes of the bush.

\subsection{Bushes of modes as dynamical objects}

	Considering bushes of modes in the previous section, we 
made use of geometrical (group-theoretical) methods only. 
Now let us consider the concrete mechanical model---the FPU-$\alpha$
  chain described by the Hamiltonian 
\begin{equation}\label{eq27}
H=\frac{1}{2}\sum_{n=1}^N{p_n^2}+\frac{1}{2}\sum_{n=1}^N{(x_{n+1}-x_n)^2}
+\frac{\alpha}{3}\sum_{n=1}^N{(x_{n+1}-x_n)^3}\,.
\end{equation}
Here $x_n(t)$ is the displacement of $n$th particle from its 
equilibrium state at the instant $t$, and $p_n(t)$ 
is the corresponding momentum.

	As in the original paper by Fermi-Pasta-Ulam \cite{l1}, the 
nonlinearity of the chain is assumed to be weak. It depends
 on two different factors --- on the coefficient $\alpha$ in 
(\ref{eq27}) and on the energy of the 
initial excitation of the system.  We can remove $\alpha$ 
from (\ref{eq27}) by means  of a scaling transformation of 
$x_n(t)$, i.e., the coefficient $\alpha$ can be made equal to~1.
Nevertheless, sometimes it is convenient to write $\alpha$
($|\alpha|\ll 1$) in Eq.~(\ref{eq27}) explicitly to stress the
weakness of the nonlinearity of the chain.

As was already stated, the basis vectors $\mbox{\boldmath$\varphi$} _j$
 (\ref{eq16})
of  irreps of the group $G_0$ are (linear) normal modes of the FPU 
chain in the harmonic approximation ($\alpha = 0$), and we 
can use them in (\ref{eq20}) as a basis for studying nonlinear 
dynamics of the system (\ref{eq27}) with $\alpha\not=0$.

We consider the equations of motion in the form,
\begin{equation}\label{eq28}
\ddot x_n=-\frac{\partial U}{\partial x_n}\;, \;\;(n=1,2,\ldots,N)\,,
\end{equation}
where $U(x_1,\ldots,x_N)$ is the potential energy of the FPU 
chain. This system of differential equations becomes linear 
in the harmonic approximation ($\alpha=0$), and reduces to 
the set of independent oscillators
\begin{equation}\label{eq29}
\ddot \mu_j + \omega_j^2\mu_j=0\;, \;\;(j=0,1,2,\ldots,N-1)
\end{equation}
with frequencies 
\begin{equation}\label{eq30}
\omega_j=2\sin\left(\frac{\pi j}{N}\right)
\end{equation}
as a result of the transition from the initial variables $x_n(t)$
to the normal modes in the complex (\ref{eq16}) or real (\ref{eq19})
form. From this point,  we call the time-dependent 
coefficients $\mu_j(t)$ [or $\nu_j(t)$] from Eq.~(\ref{eq20})
by the term, ``normal modes'', even though this term corresponds really
to $\mu_j(t)\mbox{\boldmath$\varphi$}_j$. 
This is convenient because $\mbox{\boldmath$\varphi$}_j$ 
are constant vectors and $\mu_j(t)$ are our new dynamical variables.
Substituting (\ref{eq20}) into equations (\ref{eq28}) and solving
 them with respect to variables $\mu_j(t)$, we obtain the following
 nonlinear differential equations (see details of such a transition 
in the general case in the Appendix  of Ref.~\cite{l10}):
\begin{equation}\label{eq31}
\ddot\mu_j+\omega_j^2\,\mu_j=\gamma \sum_{k=0}^{N-1}
\mu_k\;\mu_{j-k}\left[2\sin\left(\frac{2\pi}{N}(j-k)\right)-
\sin\left(\frac{2\pi}{N}j\right)\right],
\end{equation}
$$\gamma  =\frac{2\alpha i}{\sqrt N}\;,\;\;j=0,1,2,\ldots,N-1\,.$$
Hereafter, we use the following convention about indices of normal 
modes: these indices must belong to the interval $[0,\; N-1]$ and, therefore, the 
value of the mode index of the type $\pm j \pm  k$ ($j, k = 0,1,\ldots,N-1$) must be 
reduced to this interval by adding or subtracting $mN$, where $m$ is an integer 
number.

	The transition from initial variables $x_n(t)$ to normal 
modes $\nu_j(t)$ in the real form (\ref{eq19}) leads to the 
following dynamical equations for the FPU-$\alpha$ model:
\begin{eqnarray}\label{eq32}
\ddot\nu_j+\omega_j^2\,\nu_j &=& -\frac{\alpha}{\sqrt N}\sum_{k=0}^{N-1}
\nu_k\;\left[\left( \nu_{j+k} + 
\nu _{-j-k}\right)\left(2\sin\frac{2\pi}{N}k
+\sin\frac{2\pi}{N}j\right)\right. \\
 &&+ \left.\left( \nu_{j-k}-\nu _{-j+k}\right)
\left( 2\sin\frac{2\pi}{N}k - \sin\frac{2\pi}{N}j\right)
\right] \,,\;\;j=0,1,2,\ldots,N-1\,.\nonumber
\end{eqnarray}

Now we can consider the {\it dynamical} cause of the 
existence of bushes of normal modes\footnote{The 
following analysis will be based on studying equations 
(\ref{eq31}) for complex modes 
$\mu_j(t)$, but the same results can be obtained using 
equations (\ref{eq32}) for real modes $\nu_j(t)$ as well.}.
First of all, let us remember that the mode $\mu_0$
corresponds to the movement of the FPU chain as a whole 
object and, therefore, to the movement of its center of mass. 
Vibrations associated with all other modes $\mu_j(t)$, 
$(j=1,2,...,N-1)$ cannot move the center of mass because of the 
specific type of displacement patterns described by basis 
vectors $\mbox{\boldmath$\varphi $}_j$. Therefore,
the excitation of any mode $\mu_j(t)$ with $j\not=0$ cannot 
lead to the excitation of the mode $\mu_0$, as a consequence 
of the conservation of momentum. This general 
conclusion can be easy verified using the dynamical equations
(\ref{eq31}). Indeed, we obtain the following 
equation of motion for the mode $\mu_0$:
$$
\ddot\mu_0+0\mu_0 = 0\,.
$$
It is also easy to verify that mode $\mu_0(t)$ cannot excite 
any other mode, and, therefore, we can forget about 
$\mu_0(t)$ when discussing dynamical equations.

On the other hand, this mode enters into each bush of the
nonlinear chain (see, for example, Eqs.~(\ref{eq26})). What does it 
mean? We obtain bushes, as geometrical objects, from the 
idea of invariance of vibration patterns with respect 
to the subgroups of the cyclic  group $G_0$ only. But there can exist further 
causes for {\it reducing} of the number 
of modes in a given bush because of some additional symmetry properties of the 
concrete dynamical system and because of its specific character (we discuss this 
problem in Secs. 2.5 and 2.6).  In the above case of the mode $\mu_0(t)$, such 
cause is provided  by the law of conservation of momentum which is  the 
consequence of  homogeneity of the space.

	Let us assume that {\it only one} mode $\mu_6(t)$ is 
excited in the FPU-$\alpha$ chain with $N=12$  particles.
In other words, we suppose that all modes except
mode $\mu_6(t)$, are equal to zero {\it identically} 
(we call them ``sleeping modes"):
\begin{equation}\label{eq33}
\mu _j(t)\equiv 0,\; \mbox{ for} \;j\ne 6.
\end{equation}
Can such a dynamical regime exist in our nonlinear chain?
Using the above assumption, it is easy to obtain  from Eqs.~(\ref{eq31}),
\begin{eqnarray}\label{eq34}
\ddot \mu _6 + 4\mu _6\!\!\!&=&\!\!\!0,\\ 
\ddot \mu _j + \omega _j^2 \mu _j\!\!\!&=&\!\!\!0, \;\;\;
\mbox{ for all}\;\;\; j \ne 6.\nonumber
\end{eqnarray}
Therefore, the hypothesis that $\mu_6(t)\not \equiv 0$ while
$\mu_j(t)\equiv 0$ for $j \ne 6$ is self-consistent, and we have 
a pure harmonic regime $\mu _6(t)= A  \cos (\sqrt2\; t+\delta )$ 
for the FPU-$\alpha$ chain ($A, \delta$ are arbitrary 
constants). It corresponds to the one-dimensional bush 
 $\{\mbox{\boldmath$\varphi$}_6\}$ with symmetry group
 $G_1[2a]$\footnote{Let us remember once more that the mode 
$\mu _0$ is excluded from our consideration.}
 [see Eqs.~(\ref{eq26})].

	Let us consider the bush with symmetry $G_2[3a]$ 
corresponding to the second equation of (\ref{eq26}).
We must now assume that $\mu _4\not\equiv 0,\;
\mu _8\not \equiv 0$
while all other modes are equal to zero {\it identically}. 
Then, we obtain from Eqs.~(\ref{eq31})\footnote{Hereafter, we 
keep in the coefficients of the dynamical equations only
three figures after the decimal point.}:
\newcounter{letterindex}
\setcounter{letterindex}{1}
\makeatletter
\renewcommand{\@eqnnum}{(\theequation \alph{letterindex})}
\begin{eqnarray}\label{eq35}
&\;\ddot \mu_4 + 3 \;\mu_4 =-2.598\; \gamma \;\mu_8^2\,,&\\
\addtocounter{equation}{-1}
\refstepcounter{letterindex}
&\ddot \mu_8 + 3 \; \mu_8 =2.598\; \gamma \;\mu_4^2\,,&\\ 
&(\mu_j \equiv 0 \;\;\mbox{ for } j \not= 4, 8 )\,.&\nonumber
\end{eqnarray} 
\renewcommand{\@eqnnum}{(\theequation)}
\makeatother
These equations describe the two-mode dynamical regime 
where both modes  $\mu_4(t)$ and $\mu_8(t)$ (they form a conjugate pair) are 
connected with each other by the {\it force} interaction. Indeed, if the first 
oscillator, described by Eq.~(\ref{eq35}a) was excited ($\mu_4(t)\not \equiv 
0$),then the force $2.598\;\gamma\;\mu_4^2$, acting on the second oscillator, 
appears in the right-hand side (rhs) of Eq.~(\ref{eq35}b) and, 
therefore, $\mu_8(t)$ cannot be equal to zero identically 
($\mu_8(t)\not \equiv 0$). Conversely, the excitation of the mode 
$\mu_8(t)$ leads to the excitation of the mode $\mu_4(t)$ 
because of the force term $-2.598\;\gamma\;\mu_8^2$ in  
Eq.~(\ref{eq35}a). Therefore, the dynamical regimes $\mu_4(t)\not\equiv 0,
\mu_8(t) \equiv 0$ and $\mu_4(t) \equiv 0, \mu_8(t)\not \equiv 0$ cannot 
exist, and we may speak about the {\it force} interaction 
between modes $\mu_4(t)$ and $\mu_8(t)$ (for more details,
see \cite{l7,l10}).

	The existence of the {\it exact} equations (\ref{eq35}) for the 
FPU-$\alpha$ chain with $N=12$ means that there exists a
two-dimensional subspace $S_{4,8}$ of the twelve-dimensional 
configuration space spanned by the basis 
vectors $\mbox{\boldmath$\varphi $}_4$ and $\mbox{\boldmath$\varphi $}_8$ (these 
vectors are
associated with the modes $\mu_4$ and $\mu_8$) which is
 {\it dynamically}
invariant. Indeed, the trajectory corresponding to the dynamical regime 
(\ref{eq35}) in the configuration space remains in the subspace 
$S_{4,8}$ for every time $t$.

	We do not discuss the problem of finding the analytical 
solution of Eqs.~(\ref{eq35}) in the present paper, and only want 
to note that these equations describe the two-dimensional 
bush with symmetry group $G_2[3a]$ as a {\it dynamical } 
object.

	Now let us consider the bush with $G_3[4a]$ from 
Eqs.~(\ref{eq26}). Assuming that $\mu_3(t)\not\equiv 0$,  
$\mu_6(t)\not\equiv 0,\; \mu_9(t)\not\equiv 0$, and all other modes
are equal to zero, we obtain from (\ref{eq31}) the following 
dynamical system with only three degrees of freedom:
\setcounter{letterindex}{1}
\makeatletter
\renewcommand{\@eqnnum}{(\theequation \alph{letterindex})}
\begin{eqnarray}\label{eq36}
\ddot\mu_3 + 2\,\mu_3 &=& -4\,\gamma\,\mu_9\,\mu_6 \,,\\
\addtocounter{equation}{-1}
\refstepcounter{letterindex}
\ddot\mu_6 + 4\,\mu_6 &=& 2\,\gamma\,(\mu_3^2 - \mu_9^2)\,,\\
\addtocounter{equation}{-1}
\refstepcounter{letterindex}
\ddot\mu_9 + 2\,\mu_9 &=& 4\,\gamma\,\mu_3\,\mu_6 \,.
\end{eqnarray}
\renewcommand{\@eqnnum}{(\theequation)}
\makeatother

Note that modes $\mu_3$ and $\mu_9$ are conjugate. They 
have the identical symmetry groups $G_3[4a]$, while the 
mode $\mu_6$ possesses higher symmetry $G_1[2a]\supset G_3[4a]$. 
Either mode $\mu_3$ or $\mu_9$ can be the 
root mode, and $\mu_6$ is the secondary mode. Indeed, if we 
excite the mode $\mu_3(t)$ at the instant $t=0$ using the 
appropriate initial conditions, the force $2\gamma\mu_3^2$ acts
on the mode $\mu_6$ according  to Eq.~(\ref{eq36}b) and, 
therefore, this mode turns out to be excited as 
well. (We can say this in another way: Eq.~(\ref{eq36}b) cannot be 
satisfied in the opposite case $\mu_3(t)\not\equiv 0, \;
\mu_6(t)\equiv 0 \;)$.
On the other hand, it follows from Eq.~(\ref{eq36}c) that 
$\mu_9(t)$ will be also excited when $\mu_3(t)\not\equiv 0$ 
and $\mu_6(t)\not\equiv 0$  because of the force 
$4\gamma\mu_3\mu_6$  in the right-hand side of this equation. 

Thus, the excitation of the root mode
$\mu_3(t)$ brings about the excitation of the {\it 
secondary} mode $\mu_6$, but the opposite 
statement is not correct as a consequence of the specific 
asymmetric structure of bush dynamical equations with 
respect to the root and secondary modes. Indeed, the 
excitation of only  the secondary mode $\mu_6$ does not 
produce forces acting on modes $\mu_3$ and $\mu_9$,
because the forces $4\gamma\mu_6\mu_9$ and
$4\gamma\mu_6\mu_3$ are equal to zero if $\mu_3(t)\equiv 0$
and $\mu_9(t)\equiv 0$. Note that the previous geometrical 
analysis based on the consideration of the symmetry groups 
of modes (the symmetry of $\mu_6$ is higher than that of 
$\mu_3, \;\mu_9$) leads to the same conclusion which was 
obtained above by the studying of the structure of the
dynamical equations (\ref{eq36}). This is the general result proved 
by us in \cite{l10} (see Theorem 1): dynamical arguments 
must confirm the geometrical (group-theoretical) arguments\footnote{ Indeed, it 
follows as a consequence of this Theorem that when the group-theoretical method 
permits the existence of a nonzero invariant vector and, therefore, the 
existence of the secondary mode corresponding to a given root mode, there also 
exist the invariants of the parent symmetry group which provide the force terms 
in the rhs of the dynamical equations leading to the excitation of this 
secondary mode.}.
	The similar consideration of the bush with symmetry 
$G_4[6a]$ leads to the following dynamical system with five 
degrees of freedom describing the evolution of modes 
$\mu_2, \mu_4, \mu_6, \mu_8 $ and $\mu_{10}$:
\setcounter{letterindex}{1}
\makeatletter
\renewcommand{\@eqnnum}{(\theequation \alph{letterindex})}
\begin{eqnarray}\label{eq37}
\ddot\mu_2+\mu_2&=&\gamma\,[-1.732\,\mu_4\mu_{10}
-3.466\,\mu_6\mu_8\,]\,,\\
\addtocounter{equation}{-1}
\refstepcounter{letterindex}
\ddot\mu_4+3\mu_4&=&\gamma\,[\,0.866\,\mu_2^2
-2.598\,\mu_8^2-3.466\,\mu_6\mu_{10}\,]\,,\\
\addtocounter{equation}{-1}
\refstepcounter{letterindex}
\ddot\mu_6+4\mu_6&=&\!\!\!3.466\gamma\,[\,\mu_2\mu_4
-\mu_8\mu_{10}\,]\,,\\
\addtocounter{equation}{-1}
\refstepcounter{letterindex}
\ddot\mu_8+3\mu_8&=&\gamma\,[\,2.598\,\mu_4^2
-0.866\,\mu_{10}^2+3.466\,\mu_2\mu_6\,]\,,\\
\addtocounter{equation}{-1}
\refstepcounter{letterindex}
\ddot\mu_{10}+\mu_{10}&=&\gamma\,[1.732\,\mu_2\mu_8
+3.466\,\mu_4\mu_6\,]\,.
\end{eqnarray}
\renewcommand{\@eqnnum}{(\theequation)}
\makeatother

Here conjugate modes $\mu_2,\; \mu_{10}$ are root modes 
because they have the lowest symmetry
$G_4[6a]$. Conjugate modes $\mu_4,\; \mu_8$ with the 
group $G_2[3a]\supset  G_4[6a]$ and mode $\mu_6$ with 
$G_1[2a]\supset G_4[6a]$ are secondary modes. This 
geometrical fact can be again checked by the appropriate 
consideration of the structure of equations (\ref{eq37}). Indeed, the 
excitation of the mode $\mu_2(t)$ implies the appearance of 
the mode $\mu_4$ because of the force 
$0.866\,\gamma\,\mu_2^2$ in Eq.~(\ref{eq37}b), and then the 
mode $\mu_8$ appears due to the force $2.598\, \gamma \mu_4^2$ in 
Eq.~(\ref{eq37}d). The presence of modes $\mu_2,\; \mu_4,\;\mu_8$
 in the dynamical process leads to the excitation of 
the mode $\mu_6$, due to the force 
$3.466\,\gamma\,\mu_2\,\mu_4$ in Eq.~(\ref{eq37}c), and to the 
excitation of the mode $\mu_{10}$ because of the force 
$\gamma(1.732\,\mu_2\,\mu_8 + 3.466\,\mu_4\,\mu_6)$ in 
Eq.~(\ref{eq37}e). Thus, the dynamical equations (\ref{eq37})
describe the evolution of the five-dimensional bush with symmetry 
$G_4[6a]$.

The last bush from Eqs.~(\ref{eq26}) is the trivial one, because it 
contains all $12$ modes and describes the dynamics of the 
FPU-$\alpha$ chain with twelve particles in the general case.

	Thus, we found all possible bushes of modes for the case 
$N=12$ generated by the subgroups of the parent group $G_0 = T$ of pure 
translations, and  it is clear that anyone can proceed in the 
similar way for considering the case of an arbitrary $N$.

In conclusion,  let us consider 
the thermodynamical limit $N \rightarrow \infty$ for the equation of 
motion (\ref{eq32}) in the real modal space.

\subsection{Integro-differential equation for the FPU-\mbox{\boldmath$\alpha$}
chain in the limit \mbox{\boldmath$N \rightarrow\infty$}}

 As usual in the physics of crystals,
it is convenient to introduce the wave number $k$ instead 
of the mode number $j$ for the case $N\gg 1$,
\begin{equation}\label{eq37.1}
k=\frac{2\pi}{N}j\;,\;\;j=0,1,2\ldots,N-1\,.
\end{equation}
All values of $k$ belong to the interval $[\;0, 2\pi)$ which 
corresponds to the primitive cell of the reciprocal lattice 
with the period $b=\frac{2\pi}{a}$, where $a$ is a period of the 
one-dimensional lattice associated with the FPU chain in 
 equilibrium (we assume $a=1$). The density of the 
$k$-points from Eq.~(\ref{eq37.1}) increases with increasing of 
$N$. These points are distributed continuously and uniformly on 
the interval $[\;0,2\pi)$ in the limit $N\rightarrow\infty$.

	The number of terms on the right-hand side of 
each Eqs.~(\ref{eq32}) increases as $N\rightarrow\infty$, and we 
can transform the sum into a definite 
integral, assuming $\nu_k(t)$ to be a smooth function of its 
index $k$. As a result of this procedure, the following 
integro-differential equation instead of the system of 
ordinary differential equations~(\ref{eq32}) can be obtained\footnote{
Note that the coefficient $\alpha $ can be removed from Eqs.~(\ref{eq32}) as well as from 
Eq.~(\ref{eq38}) by the appropriate scale transformation of the functions $\nu_k(t)$ 
and $z(x,t)$, respectively. In Eq.~(\ref{eq38}) we also imply that the variables $\pm x$ and $\pm y$ are 
reduced to the interval $[0, 2\pi)$ with the aid of cyclic conditions,
$x\equiv x\pm 2\pi $ and $y\equiv y\pm 2\pi $.}:
\begin{eqnarray}\label{eq38}
\ddot z(x,t)+\omega^2(x)z(x,t)=\int\limits_0^{2\pi}z(y,t)
\left\{\rule{-4pt}{14pt}\left[\rule{0pt}{12pt}z(x+y,t)+z(-x-y,t)\right]
K_+(x,y)\right.\\
+\left.\left[\rule{0pt}{12pt}z(x-y,t)-z(-x+y,t)
\rule{0pt}{12pt}\right]K_-(x,y)\rule{-2pt}{14pt}\right\}dy\,, \nonumber
\end{eqnarray}
where
\begin{equation}\label{eq39}
\omega (x)=2\sin(\frac{x}{2}),\;\;K_\pm (x,y)=2\sin(y)\pm \sin(x),\;\;
0\leq x<2\pi \,.
\end{equation}
The function $z(x,t)$ was introduced in Eq.~(\ref{eq38}) instead 
of $\nu_k(t)$ and, therefore, $z(x,t)$ signifies  the 
value of the mode with wave number $x$ at time $t$.
Thus, we use {\it different} notations for the wave number in 
the discrete case $(k)$ and in the continuous case $(x)$. It is
easy to verify the correctness of Eqs.~(\ref{eq38}, \ref{eq39}) by replacing the 
integral on the right-hand side of Eq.~(\ref{eq38}) with its Darboux 
sum and using the following relation 
between $z(x,t)$ and $\nu_k(t)$:
\begin{equation}\label{eq40}
z(x,t) = \frac{\sqrt N}{2\pi }\;\nu_k(t)\,,\;\; (x=k).
\end{equation}
Such a procedure returns us to the discrete model (\ref{eq32}).

Note that Eqs.~(\ref{eq38}, \ref{eq39}) correspond to the continuum limit 
for the FPU-$\alpha$ model in {\it modal} space, while 
the well-known Korteweg-deVries equation (KdV) 
corresponds to the continuum limit  in {\it ordinary} 
space \cite{l4}. Besides the form of above mentioned  
equations, there exists a principal difference  between these 
two equations. Indeed, Eqs.~(\ref{eq38}, \ref{eq39}),
 in contrast to the KdV-equation, 
is valid not only  for long waves, but for short 
waves as well. We can make use  of this fact for considering 
bushes of modes which often correspond namely to 
sufficiently short waves. The application of the integro-differential 
equation (\ref{eq38}) for studying some problems of the 
FPU-$\alpha$ dynamics will be considered elsewhere.

\subsection{On the difference between bush structures for FPU-$\alpha$ and 
\mbox{FPU-$\beta$} chains} 

	Let us compare our results, which are based on the group-theoretical
 method (Sec. 2.2), with those by Poggi and Ruffo 
\cite{l3}, which are based on specific dynamical properties  
of the FPU-$\beta$ chain. Naturally, taking into account 
concrete properties  of a given mechanical system, we can 
obtain some {\it additional} selection rules as compared
with general symmetry-related results.
	Poggi and Ruffo found for the FPU-$\beta$ model five 
one-mode solutions corresponding to the following numbers of 
modes
\begin{equation}\label{eq48}
j=\frac{\raisebox{-2pt}{N}}{4}\,,\;\frac{\raisebox{-2pt}{N}}{3}\,,\;
\frac{\raisebox{-2pt}{N}}{2}\,,\;\frac{\raisebox{-2pt}{2N}}{3}\,,\;
\frac{\raisebox{-2pt}{3N}}{4}
\end{equation}
and two types of solutions associated with the pairs of modes
   $\{\frac{N}{4},\frac{3N}{4}\}$ and $\{\frac{N}{3},\frac{2N}{3}\}$. 
Our results differ from 
these, and we want to discuss the problem using the example of 
the chain with $N=12$ considered in Sec. 2.2.

According to Eqs. (\ref{eq26}) there exists only one  one-dimensional bush 
$B[2a]$ associated with the excitation  of the mode with 
$j=\frac{N}{2}=6$ (remember that we, as well as Poggi and Ruffo, neglect 
the mode $\mu_0$). But the excitation of the mode with 
$j=\frac{N}{3}=4$ leads to the appearance of the two-dimensional 
bush $B[3a]$ with $j=4,8$ (modes $\mu_4$ and $\mu_8$ are 
conjugate). Thus, in contrast to the FPU-$\beta$ model, 
the one-dimensional dynamical regime with {\it only} $\mu_4$ (or $\mu_8$)
does not exist. Moreover, this conclusion is confirmed also 
by the {\it dynamical} consideration for the FPU-$\alpha$ 
model in the Sec. 2.3. The difference between structures of the 
bush $B[3a]$ for the FPU-$\beta$ and for the FPU-$\alpha$ 
chains is brought about namely by their specific dynamical 
properties.

Because of the same reason, the excitation of the mode 
$\mu_3 (j=\frac{N}{4})$ leads, for the FPU-$\alpha$ chain and for the 
general case as well, to the excitation of the conjugate mode $\mu_9 
(j=\frac{3N}{4})$ and to the secondary mode $\mu_6 (j=\frac{N}{2})$.
 We see, therefore, that some modes in the list (\ref{eq48}),
 obtained by Poggi and Ruffo 
for the FPU-$\beta$ chain, combine into the three-dimensional 
bush $B[4a]$ for the FPU-$\alpha$ model, as well as for the 
arbitrary nonlinear chain. 
In other words, the three-dimensional symmetry-determined bush
$B[4a]$, being an indivisible nonlinear dynamical object for FPU-$\alpha $
model, can be split into {\it smaller} invariant manifolds (i.e., bushes of modes 
in general sense) by imposing the appropriate initial conditions. Indeed, the 
one-dimensional bush with $j= \frac{N}{4}$ can exist in the FPU-$\beta $ model
according to Eq.~(\ref{eq48}), but it does not exist in the FPU-$\alpha $ model
because it follows from Eq.~(\ref{eq36}b) that excitation of the mode $\mu _3 (j= \frac{N}{4})$
{\it always} leads to the excitation of the (secondary) mode $\mu _6(j= \frac{N}{2})$.

	Let us stress once again that our results, obtained by the 
group-theoretical method, are general: they are correct for any 
nonlinear monatomic chain\footnote{In some sense, they are also valid for 
multiatomic chains (see Sec.~2.6).}. Nevertheless, the specific 
character of interatomic interactions can {\it reduce}  the 
dimensionality of bushes found by this method, as we have just 
seen comparing the structures of bushes of modes for 
FPU-$\alpha$ and FPU-$\beta$ chains.

\subsection {Bushes of normal modes associated with dihedral symmetry of 
nonlinear chains}

Introducing the concept of bushes of normal modes, we started with a certain 
symmetry group $G_0$ (parent group) of the considered mechanical system and basis vectors 
of its irreducible representations. Each given bush corresponds to some 
subgroup $G_k$ of the group $G_0 (G_k \subset  G_0)$.

In previous sections, the group $T={E, \hat a, \hat a^2, \ldots, \hat a^{N-1}}$ (see Eq.(4)) was 
chosen as the symmetry group $G_0$ of the monatomic chain in equilibrium. 
But the {\it full} symmetry of such a system is higher, because space inversion 
$\hat i$ with respect to the center of the finite monatomic chain also is a 
symmetry element, as well as the product $\hat a^k \cdot \hat  i$ of the translation 
$\hat a^k (k=0, 1, \ldots, N-1)$ with the inversion $\hat i$.
 Therefore, the full symmetry group $D$ (dihedral group) of the monatomic 
chain possesses twice as many elements as the translational group 
$T$ which is its subgroup of index $2$. We can write the group $D$ as the direct 
sum of two {\it cosets} with respect to the subgroup $T$:
\begin{equation} \label{eq100}
D = T + T\cdot \hat i.                                                 		
\end{equation}
It is convenient for our purpose to rewrite this equation in the 
form
\begin{equation} \label{eq101}
\tilde G_0 = G_0 + G_0\cdot \hat i,		
\end{equation}
where $G_0=T$ and $\tilde G_0 =D$.

We will discuss the connection between bushes of modes associated with the 
group $G_0$ and those associated with its supergroup $\tilde G_0$.
The irreducible representations of the groups $G_0$ and $\tilde G_0$ will
be denoted by
$\Gamma_j$ and $\tilde \Gamma_j$, respectively.

Up to this point we discussed bushes for monatomic chains in terms of the 
irreps $\Gamma_j$ of the translational group $G_0=T$, although the full 
symmetry group of these systems is $\tilde G_0=D$. 
Why did we proceed in such a manner and how is the list of bushes of normal 
modes modified if we consider the complete symmetry group 
$\tilde G_0$ instead of its subgroup $G_0$? 
We give the brief answer to this question below, while the detailed consideration 
will be given in another paper devoted to multiatomic chains.

The group $\tilde G_0=D$ is a non-Abelian group since some its elements do not 
commute with each other (for example, $\hat i\cdot \hat a = \hat a^{-1}\cdot \hat i$).
 As a consequence, the number of classes of conjugate elements of this group is 
less than the total number ($2N$) of its elements and some irreps $\tilde \Gamma_j$ 
are not one-dimensional.

Let us discuss the monatomic chain with $N=12$ particles in more detail.
The appropriate group $\tilde G_0=D$ has nine classes of conjugate elements 
and, therefore, nine  irreducible representations $\tilde \Gamma_j$.
 Four of these irreps ($\tilde \Gamma _1, \tilde \Gamma _2, \tilde \Gamma _3, \tilde \Gamma _4$)
 are one-dimensional and five other irreps ($\tilde \Gamma _5, \tilde \Gamma _6,
 \tilde \Gamma _7, \tilde \Gamma _8, \tilde \Gamma _9$) are two-dimensional. 
	
To determine a given irrep $\tilde \Gamma_j$ of the group $\tilde G_0$ defined
 by Eq.~(\ref{eq101}), we must write matrices of this representation for
 elements of the first coset $G_0$ and of the second coset $G_0\cdot \hat i$.
 They form the first half $\tilde \Gamma ^{(1)}_j$ and the second half
 $\tilde \Gamma ^{(2)}_j$ of the irrep $\tilde \Gamma _j$, 
respectively.\footnote
{Obviously, $\tilde \Gamma _j^{(1)}$ is the {\it restriction} of
 the irrep $\tilde \Gamma _j$ of the group $\tilde G_0$ onto the subgroup $G_0$ .}
$\tilde \Gamma ^{(1)}_j$ for one-dimensional irreps coincides
 with either irrep $\Gamma _0$ or irrep $\Gamma _6$ of the 
group $G_0$ (see Eq.~(\ref{eq103}) below). $\tilde \Gamma _j^{(1)}$ for
 two-dimensional irreps can be written as a direct 
sum of two {\it conjugate} irreps of group $G_0$, for example,
 $\tilde \Gamma _5^{(1)}=\pmatrix{\Gamma _1& \cr  &\Gamma _{11}}$.
 All matrices of the second half $\tilde \Gamma _j^{(2)}$ of the irrep $\tilde \Gamma _j$
 are products of the appropriate matrices of $\tilde \Gamma _j^{(1)}$ with 
the matrix corresponding to the inversion $(\hat i)$.
 For one-dimensional irreps $\tilde \Gamma _j$  this 
matrix is equal to (+1) or (-1), while for two-dimensional irreps $\tilde \Gamma _j$
 it is equal to $\pmatrix{0&1 \cr 1 &0}$.
Therefore, the irrep $\tilde \Gamma _5$ has the following form\footnote
{Henceforth, we will omit zero elements of all matrices.}
\begin{equation} \label{eq102}
\tilde \Gamma _5 =	\pmatrix{\Gamma _1& \cr  &\Gamma _{11}} +
\pmatrix{\Gamma _1& \cr  &\Gamma _{11}}	\pmatrix{&1 \cr 1 &}.
\end{equation}
 This view of the irrep $\tilde \Gamma _j$ must be compared with the 
definition (\ref{eq101}) of group $\tilde G_0$;
it allows one to determine the explicit form of all matrices of the both cosets 
from Eq.~ (\ref{eq101}).

 Using the above notation we can write all irreps $\tilde \Gamma _j \; (j = 1,..., 9)$ 
 of the group $\tilde G_0$ as follows:
\begin{eqnarray} \label{eq103}
&&\tilde \Gamma _1 =	\Gamma _0 + \Gamma _0 \cdot  (1);\;\;\;
\tilde \Gamma _2 =	\Gamma _0 + \Gamma _0 \cdot  (-1);\nonumber\\	
&&\tilde \Gamma _3 =	\Gamma _6 + \Gamma _6 \cdot  (1);\;\;\;
\tilde \Gamma _4 =	\Gamma _6 + \Gamma _6 \cdot  (-1);\\
&&\tilde \Gamma _k =	 \pmatrix{\Gamma _{k-4}& \cr  &\Gamma _{16-k}} +
\pmatrix{\Gamma _{k-4}& \cr  &\Gamma _{16-k}}	\pmatrix{&1 \cr 1 &}
,\;\;k= 5, 6, 7, 8, 9. \nonumber
\end{eqnarray}
Note that irreps $\Gamma _{k-4}$ and $\Gamma _{16-k}$ in Eq.~(\ref{eq103})
 are mutually conjugate for any $k$: 
\begin{equation} \label{eq104_1}
 \Gamma _1 - \Gamma _{11},\;\;
 \Gamma _2 - \Gamma _{10},\;\;
 \Gamma _3 - \Gamma _{9},\;\;
 \Gamma _4 - \Gamma _{8},\;\;
 \Gamma _5 - \Gamma _{7}.\;\;
\end{equation}

As was already discussed, bushes of normal modes correspond to different subgroups of the parent symmetry 
group of the considered mechanical system in equilibrium. For monatomic chains, 
bushes associated 
with the parent group $G_0$ were considered in Sec. 2.2. Now we want to discuss 
bushes associated with the parent group $\tilde G_0$.

There are 34 subgroups of the group $\tilde G_0$ for the chain with $N=12$ particles.
Note that since $G_0\subset \tilde G_0$, all subgroups 
of $G_0$ are also subgroups of $\tilde G_0$. 
Therefore, all bushes found in Sec. 2.2 are valid not only for the parent group 
$G_0$, but for the parent group $\tilde G_0$, too.
 In addition to these bushes, some {\it new} bushes of normal modes can appear 
when we consider $\tilde G_0$ as the parent group. To find them we must 
examine those subgroups of $\tilde G_0$ which contain some elements of the 
second coset ($G_0\cdot \hat i$) from Eq. (\ref{eq101}).
 Obviously, all these new subgroups can be 
constructed by adding certain elements of $G_0\cdot \hat i$ to the 
subgroups of the group $G_0$ (in particular, to the trivial subgroup \{$E$\}).
Let us consider a concrete example.

 The subgroup \{$\hat a^4, \hat i$\} of the group $\tilde G_0$ is determined 
by two generators $\hat a^4$ and $\hat i$ (hereafter we write down
 in braces {\it only generators} of the appropriate group instead of all its elements). 
This subgroup is obtained from 
the subgroup \{$\hat a^4$\} of the group $G_0=T$ by adding inversion $\hat i$. As was 
shown in Sec. 2.2, the three-dimensional vibrational bush 
\begin{equation} \label{eq104}
B\{\hat a^4\}=\mu _6(t)  \mbox{\boldmath$\varphi $}_6 + \mu _3(t)  \mbox{\boldmath$\varphi $}_3
+\mu _9(t)  \mbox{\boldmath$\varphi $}_9      
\end{equation}
corresponds to the subgroup \{$\hat a^4$\} where three functions $\mu _6(t)$,
 $\mu _3(t)$ and $\mu _9(t)$ are {\it independent} of each other. 
Thus, three one-dimensional irreps $\Gamma_6$, $\Gamma_3$
and $\Gamma _9$ contribute to the bush $B \{\hat a^4 \}$ associated with 
the parent group $G_0 = T$ (namely 
their basis vectors $\mbox{\boldmath$\varphi $}_j$ are contained in Eq. (\ref{eq104}).

In terms of irreps of the parent group $\tilde G_0$, the {\it same} bush is formed by 
basis vectors of one-dimensional irrep $\tilde \Gamma_3$ and two-dimensional irrep 
$\tilde \Gamma_7$ (note that restrictions of these irreps to the subgroup \{$\hat a^4$\} 
give us the above mentioned irreps $\Gamma_6$, $\Gamma_3$ and $\Gamma_9$ 
of the group $G_0=T$). The additional generator $\hat i$ which appears when 
we pass from the subgroup \{$\hat a ^4$\} to the subgroup \{$\hat a^4,\;\hat i$\}
 introduces the matrix (1) for the irrep $\tilde \Gamma_3$ and the matrix 
$\pmatrix{&1 \cr  1&}$ for the irrep $\tilde \Gamma_7$, and the new bush $B\{\hat a, \hat i\}$
 must be invariant with respect to these matrices. This 
invariance condition leads to the following transformation under the action of $\hat i$: 
\begin{equation} \label{eq105}
\mu _6(t)\rightarrow \mu _6(t),\;\mu _3(t)\rightarrow \mu _9(t),\;
\mu _9(t)\rightarrow \mu _3(t)\;.       
\end{equation}
As a result, we obtain the restriction $\mu _3(t)=\mu _9(t)$ on previously
 independent functions $\mu _3(t)$ and $\mu _9(t)$.
 In other words, the old three-dimensional bush $B\{\hat a^4\}$ from Eq. (\ref{eq104}) 
transforms into the new two-dimensional bush   
\begin{equation} \label{eq106}
B\{\hat a^4,\;\hat i\}=\mu _6(t)  \mbox{\boldmath$\varphi $}_6 + \mu _3(t)
[ \mbox{\boldmath$\varphi $}_3 +  \mbox{\boldmath$\varphi $}_9 ]     
\end{equation}	

 Let us emphasize that the transition from the parent group $G_0=T$ to the parent 
group $\tilde G_0 = D$ results in {\it  new superposition's} of one and the 
{\it same} set of normal modes $\mbox{\boldmath$\varphi $}_j\;(j = 0, 1,\ldots, 11)$
 when we find the appropriate bushes of modes and, therefore, we can 
compare bushes associated with different parent groups. 

 Thus, all bushes associated with the group $G_0$ are also bushes associated with 
the group $\tilde G_0$, but there appears a set of new 
bushes for $\tilde G_0$ because of additional restrictions
on time-dependent functions $\mu _j(t)$ for some old bushes.
The detailed analysis of this phenomenon will be published elsewhere, 
while we give the appropriate results in Table~2 of the present paper. 

The dimensions of the bushes are indicated in the first column of this table.
The list of the appropriate irreps $\Gamma _j$ in square brackets 
in column 2 corresponds to a certain bush. Such a
list determines the bush uniquely, since a quite definite basis
 vector $\mbox{\boldmath$\varphi $}_j$ corresponds to the irrep
 $\Gamma _j$ of the group $G_0$.

For bushes associated with the parent group $\tilde G_0 = D$ (column 3) 
we give in square brackets not only individual irreps $\Gamma _j$
 (of the group $G_0 = T$!) but also the {\it pairs} $\Gamma _i -\Gamma _j$ of 
conjugate irreps (the irreps of such pairs are connected by hyphens). 
{\it Only one} function $\mu _i(t)$ corresponds to each pair $\Gamma _i -\Gamma _j$. 
The above discussed bushes (\ref{eq104}) and (\ref{eq106}) can 
be written in this notation as follows: 
\begin{equation} \label{eq107}
B\{\hat a^4\}=[\Gamma _6, \Gamma _3, \Gamma _4],
\end{equation}	    
\begin{equation} \label{eq108}
B\{\hat a^4,\;\hat i\}=[\Gamma _6, \Gamma _3 - \Gamma _4].
\end{equation}	

The relation between $\mu _3(t)$ and $\mu _9(t)$ for the 
bush $B\{\hat a^4, \hat i\}$ (see Eq. (\ref{eq106})) is very simple 
($\mu _3(t) = \mu _9(t)$), while for some other bushes the similar relations
 can be more complicated. Namely because of this reason we do not
 define concretely the relation between functions $\mu _i(t)$ and $\mu _j(t)$, 
and write simply $\Gamma _i - \Gamma _j$, 
indicating by the same token only that such a relation does exist and, therefore,
instead of two independent functions $\mu _i(t)$ and $\mu _j(t)$ only one 
function $\mu _i(t)$ corresponds to the pair $\Gamma _i - \Gamma _j$.

 Moreover, each square brackets in the third column of Table 3 can correspond to 
{\it several} bushes which differ from each other by different relations between 
$\mu _i(t)$ and $\mu _j(t)$ associated with pairs $\Gamma _i - \Gamma _j$. 
By the way, this is one of the causes why the 
number of bushes in Table 3 is less than the total number (34) of subgroups of the 
group $\tilde G_0 = D$. 
There exist also another cause of the above phenomenon. Indeed, there are no 
{\it vibrational} bushes for some subgroups of the parent 
group $\tilde G_0 = D$.
As examples we can point to subgroups \{$\hat a, \hat i$\} and \{$\hat a^2, \hat a\hat i$\}.
Finally, it is possible that 
adding a new generator does not change the appropriate old bush (for 
example, $B \{\hat a^2, \hat i\} = B \{\hat a^2\}$).

To conclude the discussion of relations between bushes associated 
with parent groups $G_0$ and $\tilde G_0$, which are indicated in Table~2, 
let us say that besides five 
bushes associated with $G_0$ there exist seven {\it new types}
of bushes in accordance with their classification 
under the parent group $\tilde G_0$: 2 one-dimensional, 2 two-dimensional, 
1 three-dimensional, 1 five-dimensional and 1 six-dimensional.

 Let us stress that for some {\it multiatomic} chains inversion is not a symmetry 
element\footnote{In terms of crystallography, we can say that it is possible for a given 
crystal to not have inversion even though its Bravais lattice does.} and $G_0 = T$ is the full 
symmetry group of these mechanical systems. In such cases we only need to use 
$G_0=T$ as the parent group for finding normal modes.
 For monatomic chains and for multiatomic chains with inversion, using 
$G_0$ as a parent group provides more rich information about possible bushes 
and, therefore, is preferable.
In most sections of the present paper, we restrict ourselves to studying only
bushes associated with the group $G_0=T$, since this case is more general and 
more simple.

It is very essential that our list of bushes given in Sec.~2.2 (see also the second 
column of Table~2) is valid for {\it every} one-dimensional structure with the 
symmetry group $G_0=T$, {\it independently} of how many particles are in its 
primitive cell: the difference appears only in the {\it form of basis vectors}, 
but not in the bush structure.
Let us discuss this point using as an example the 
bush with subgroup $G_3 [4a]$ from Eq.~(\ref{eq26}) which was denoted in the 
present section as $B \{\hat a^4\}$ (Eq.~(\ref{eq104})).

The bush of such type can be excited in various chains, but the basis vectors 
$ \mbox{\boldmath$\varphi $}_j$ from (\ref{eq104}) will be different for
different structures of the primitive cell. In particular,
the dimension of the vectors $ \mbox{\boldmath$\varphi $}_j$ and 
$ \mbox{\boldmath$X$}(t)$ is equal to $k\cdot  N$ if there are
$k$ particles in the primitive cell. 
Moreover, the {\it dimension} of the bush also can change, because, in general, 
basis vectors $ \mbox{\boldmath$\varphi $}_j$ depend on a number of 
arbitrary constants and this number is equal to the number of copies of the
appropriate irreps in the decomposition of the full vibrational representation of 
our mechanical system into its irreducible parts [10].

Thus, the group-theoretical methods give us a possibility to divide the 
problem of finding all types of symmetry-determined nonlinear vibrations into 
two independent parts:

1) obtaining the list of bushes using group-subgroup relations only;

2) taking into account the concrete structure of the considered mechanical 
systems.

We will discuss bushes of normal modes for multiatomic chains in more detail 
in a special paper.

\section{Stability of bushes of normal modes}
Some aspects of the stability of exact solutions 
for the FPU-$\beta$ chain
were discussed by Poggi and Ruffo in \cite{l3}. They used the 
standard method based on the idea of the linearization of appropriate 
dynamical equations in the vicinity of a given solution.
We also make use of this method, as well as of a direct numerical study
of the dynamical equations for the FPU-$\alpha$ chain in the ordinary
and in the modal space.

 The loss of the stability of bushes for small $N$ and for large $N$ can occur in different ways.
It is convenient to discuss these cases separately.

\subsection{Stability of bushes of modes for small $N$}
Let us consider the problem of bush stability using as an example 
the FPU-$\alpha$ chain with $N=12$.
All numerical data on thresholds of the bush stability, given in
the present section, correspond namely to this case.

 Firstly, we discuss the stability 
of the one-dimensional bush $B[2a]$ with the symmetry group $G=[2a]$. 
It contains only one mode $\nu_6(t)$. Therefore, we must consider all 
other modes $(j \not =6)$ in Eqs. (\ref{eq32}) to be equal to 
zero in the exact solution corresponding to the given bush. 
Bearing in mind the intention to study the bush stability, 
we suppose that these ``sleeping" modes are not exactly equal to zero, 
but are sufficiently small quantities (for example, at about $10^{-10}$ 
for the numeric investigation). Then the terms of Eqs. (\ref{eq32}) 
differ from each other by the order of their smallness:
\begin{equation}\label{eq49}
\mid\nu_j\nu_6\mid\;\;\gg\;\mid\nu_i\nu_k\mid \;\;
\mbox{for}\;i, j, k \not = 6 \,.
\end{equation}
As a result, we can keep only those terms in the 
right-hand sides of Eqs.~(\ref{eq32}), which contain the 
mode $\nu_6(t)$. 
Thus, these equations can be reduced for $N=12$ to the following form
(hereafter all numerical coefficients are given up to
 the third figure after the decimal point):
\setcounter{letterindex}{1}
\makeatletter
\renewcommand{\@eqnnum}{(\theequation \alph{letterindex})}
\begin{eqnarray}\label{eq50}
\ddot\nu_1+\omega_1^2\nu_1&=&-1.155\;\alpha\,\nu_5\nu_6\,,\\
\addtocounter{equation}{-1}
\refstepcounter{letterindex}
\ddot\nu_2+\omega_2^2\nu_2&=&-2\;\alpha\,\nu_4\nu_6\,,\\
\addtocounter{equation}{-1}
\refstepcounter{letterindex}
\ddot\nu_3+\omega_3^2\nu_3&=&-2.309\;\alpha\,\nu_3\nu_6\,,\\
\addtocounter{equation}{-1}
\refstepcounter{letterindex}
\ddot\nu_4+\omega_4^2\nu_4&=&-2\;\alpha\,\nu_2\nu_6\,,\\
\addtocounter{equation}{-1}
\refstepcounter{letterindex}
\ddot\nu_5+\omega_5^2\nu_5&=&-1.155\;\alpha\,\nu_1\nu_6\,,\\
\addtocounter{equation}{-1}
\refstepcounter{letterindex}
\ddot\nu_6+\omega_6^2\nu_6&=&0\,,\\
\addtocounter{equation}{-1}
\refstepcounter{letterindex}
\ddot\nu_7+\omega_7^2\nu_7&=&1.155\;\alpha\,\nu_{11}\nu_6\,,\\
\addtocounter{equation}{-1}
\refstepcounter{letterindex}
\ddot\nu_8+\omega_8^2\nu_8&=&2\;\alpha\,\nu_{10}\nu_6\,,\\
\addtocounter{equation}{-1}
\refstepcounter{letterindex}
\ddot\nu_9+\omega_9^2\nu_9&=&2.309\;\alpha\,\nu_9\nu_6\,,\\
\addtocounter{equation}{-1}
\refstepcounter{letterindex}
\ddot\nu_{10}+\omega_{10}^2\nu_{10}&=&2\;\alpha\,\nu_8\nu_6\,,\\
\addtocounter{equation}{-1}
\refstepcounter{letterindex}
\ddot\nu_{11}+\omega_{11}^2\nu_{11}&=&1.155\;\alpha\,\nu_7\nu_6\,,
\end{eqnarray}
\renewcommand{\@eqnnum}{(\theequation)}
\makeatother
where
\begin{equation}\label{eq51}
\omega_j^2=4\sin^2\left(\frac{\pi j}{12}\right)\;,\;\;
j=1,2,\ldots ,11\,.
\end{equation}
The system (\ref{eq50}) of eleven equations splits into seven 
independent subsystems. Moreover, it follows from the equation (\ref{eq50}f)\footnote{     
Let us emphasize that this is the {\it exact} equation of the one-dimensional 
bush $B[2a]$ (see also Eq.~(\ref{eq34})).}
\begin{equation}\label{eq52}
\ddot\nu_6+\omega_6^2\nu_6=0\;\;\;\mbox{$(\omega_6^2=4)$}
\end{equation}
that the root mode $\nu_6(t)$ vibrates in harmonic manner  ,
\begin{equation}\label{eq53}
\nu_6(t)=A\cos(2t+\delta),
\end{equation}
and, therefore, we can substitute $\nu_6(t)$ in this 
form into all of the other equations in (\ref{eq50}). 
Thus, we obtain
\setcounter{letterindex}{1}
\makeatletter
\renewcommand{\@eqnnum}{(\theequation \alph{letterindex})}
\begin{eqnarray}\label{eq54}
\ddot\nu_1+\omega_1^2\nu_1&=&-1.155(\alpha A)\nu_5\cos(2\tau)\,,\\
\ddot\nu_5+\omega_5^2\nu_5&=&-1.155(\alpha A)\nu_1\cos(2\tau)\,,\nonumber\\
\nonumber\\
\addtocounter{equation}{-1}
\refstepcounter{letterindex}
\ddot\nu_2+\omega_2^2\nu_2&=&-2(\alpha A)\nu_4\cos(2\tau)\,,\\
\ddot\nu_4+\omega_4^2\nu_4&=&-2(\alpha A)\nu_2\cos(2\tau)\,,\nonumber\\
\nonumber\\
\addtocounter{equation}{-1}
\refstepcounter{letterindex}
\ddot\nu_3+\omega_3^2\nu_3&=&-2.309(\alpha A)\nu_3\cos(2\tau)\,,\\
\nonumber\\
\addtocounter{equation}{-1}
\refstepcounter{letterindex}
\ddot\nu_7+\omega_7^2\nu_7&=&1.155(\alpha A)\nu_{11}\cos(2\tau)\,,\\
\ddot\nu_{11}+\omega_{11}^2\nu_{11}&=&1.155(\alpha 
A)\nu_7\cos(2\tau)\,,\nonumber\\
\nonumber\\
\addtocounter{equation}{-1}
\refstepcounter{letterindex}
\ddot\nu_8+\omega_8^2\nu_8&=&2(\alpha A)\nu_{10}\cos(2\tau)\,,\\
\ddot\nu_{10}+\omega_{10}^2\nu_{10}&=&2(\alpha A)\nu_8\cos(2\tau)\,,\nonumber\\
\nonumber\\
\addtocounter{equation}{-1}
\refstepcounter{letterindex}
\ddot\nu_9+\omega_9^2\nu_9&=&2.309(\alpha A)\nu_9\cos(2\tau)\,.
\end{eqnarray}
\renewcommand{\@eqnnum}{(\theequation)}\makeatother
The initial phase $\delta$ was removed from the
above equations by introducing the new time variable
\begin{equation}\label{eq55}
\tau=t+\frac{\delta}{2}\,.
\end{equation}
Individual subsystems (\ref{eq54}) describe different ways that the
stability of the original bush $B[2a]$ can be lost. Indeed, we obtain 
conditions of the loss of stability with respect to the modes 
$\nu_3(t)$ and $\nu_9(t)$ from Eqs (\ref{eq54}c) and (\ref{eq54}f),
respectively,  which can be easily reduced to the standard form of the
Mathieu equation  \cite{l19}:
\begin{equation}\label{eq56}
\ddot y+[a-2q\cos(2\tau)]y=0\,.
\end{equation}
The set of stable and unstable regions correspond to this
equation in the plane $(a,q)$ of its pertinent parameters.
Using the stability chart of the Mathieu equation in the 
same manner as in \cite{l3} we obtain, for the considered way 
of the loss of stability, that the 
bush $B[2a]$, describing by the mode $\nu _6(t)$, is stable for
\begin{equation}\label{eq56*}
\mid\alpha A\mid\;<(\alpha A)_c = 1.049\,.
\end{equation}

Moreover,  it can be stable for larger values of 
$\mid\alpha A\mid$, as well (for example, for the interval 
$16.465<\mid\alpha A\mid<16.474$), but in this paper we study
only the first (basic) zone of stability for all bushes of modes.

Appearance of the modes $\nu_3 $ and $\nu_9 $ means that the original
one-dimensional bush $B[2a]$ enlarges up to the three-dimensional bush 
$B[4a]$ which embraces the root modes $\nu_3$,$\nu_9$ and the secondary
mode $\nu_6$ [see Eq.~(\ref{eq26})].
Let us rewrite Eqs.~(\ref{eq50}c), (\ref{eq50}f) and 
(\ref{eq50}i), describing the dynamics of this bush:
\setcounter{letterindex}{1}
\makeatletter
\renewcommand{\@eqnnum}{(\theequation \alph{letterindex})}
\begin{eqnarray}\label{eq57}
\ddot\nu_3+2\nu_3&=&-2.309\;\alpha\;\nu_3\nu_6\,,\\
\addtocounter{equation}{-1}
\refstepcounter{letterindex}
\ddot\nu_6+4\nu_6&=&0\,,\\
\addtocounter{equation}{-1}
\refstepcounter{letterindex}
\ddot\nu_9+2\nu_9&=&2.309\;\alpha\;\nu_9\nu_6\,.
\end{eqnarray}
\renewcommand{\@eqnnum}{(\theequation)}\makeatother
Eqs. (\ref{eq57}) represent the linearized equations
(with respect to the ``sleeping" modes $\nu_3$ and $\nu_6$)
of the {\it exact} equations of the bush $B[4a]$ in terms of real
modes\footnote{Compare these equations with those in terms of complex 
modes (\ref{eq36}).}:           
\setcounter{letterindex}{1}
\makeatletter
\renewcommand{\@eqnnum}{(\theequation \alph{letterindex})}
\begin{eqnarray}\label{eq159}
\ddot\nu_3+ \omega _3^2 \nu _3&=&-\frac {8\alpha}{\sqrt{12}}\;\nu_3\nu_6\,,\\
\addtocounter{equation}{-1}
\refstepcounter{letterindex}
\ddot\nu_6+ \omega _6^2 \nu _6&=&-\frac {4\alpha}{\sqrt{12}}\;(\nu _3^2 - \nu _9^2)\,,\\ 
\addtocounter{equation}{-1}
\refstepcounter{letterindex}
\ddot\nu_9+ \omega _9^2 \nu _9&=&\frac {8\alpha}{\sqrt{12}}\;\nu_9\nu_6\,. \label{eq159ñ}
\end{eqnarray}
\renewcommand{\@eqnnum}{(\theequation)}
\makeatother
In contrast to the equations (\ref{eq36})
for the bush $B[4a]$ in terms of complex modes
$\mu_3, \mu_6, \mu_9$, Eqs.~(\ref{eq159}a) and (\ref{eq159}c) are
independent from each other. Nevertheless, the parametric excitation
of the mode $\nu_3(t)$, brought about by the mode 
$\nu_6(t)=A\cos(2t+\delta)$, leads simultaneously to the 
excitation of the mode $\nu_9(t)$. Indeed, the same condition of the 
loss of the stability corresponds to both equations (\ref{eq57}a) 
and (\ref{eq57}c), because they can be converted to the identical 
form using the transformation (\ref{eq55}) with the initial phase 
shifted by $\pi$.

Equations (\ref{eq54}b) for modes $\nu_2(t)$, $\nu_4(t)$ and 
(\ref{eq54}e) for conjugate modes $\nu_{10}(t)$, $\nu_8(t)$ describe 
the transition from the original bush $B[2a]$ to the embracing bush 
$B[6a]$ because of the parametric resonance with the mode $\nu_6(t)$
[see (\ref{eq53})]. Note that Eqs. (\ref{eq54}b) and (\ref{eq54}e) 
can be converted to the same form using the above mentioned 
transformation reducing Eqs. (\ref{eq57}c) to the form of 
Eqs (\ref{eq57}a) and, therefore, we may discuss equations for 
modes $\nu_2(t)$ and $\nu_4(t)$ only.

	It is interesting to stress that we cannot consider the parametric 
excitation of the bush $B[3a]$ by the mode $\nu_6(t)$ belonging to the 
bush $B[2a]$. 
Indeed, according to Eq.(\ref{eq26}) the two-dimensional vibrational bush $B[3a]$
consist of two modes, $\nu _4$ and $\nu _8$, whose symmetry 
group\footnote{See Table 1.} is $G = 3a$. On the other hand, 
the new bush $B$, appearing as a result of the loss of stability
of the old bush $B[2a]$, must {\it include} the last bush and, therefore,
must contain the mode $\nu_6$.
Thus, the above mentioned loss of stability can induce only bushes 
$B[6a]$ and $B[12a]$, but not the bush $B[3a]$ (see Eq.~(\ref{eq26})).
This geometrical fact can be confirmed by the following 
dynamical arguments.

We find from Eqs.~(\ref{eq54}) that the active mode $\nu_6(t)$ (see Eq.~(\ref{eq53}))
can excite the following {\it pairs} of the sleeping modes $\nu_1$, $\nu_5$ 
(Eq.~(\ref{eq54}a)), $\nu_2$, $\nu_4$ (Eq.~(\ref{eq54}b)),
$\nu_7$, $\nu_{11}$ (Eq.~(\ref{eq54}d))
and  $\nu_8$, $\nu_{10}$ (Eq.~(\ref{eq54}e)).
All modes in the pairs ($\nu_1$, $\nu_5$) and ($\nu_7$, $\nu_{11}$)
possess the same symmetry group $G = 12$, while the modes from each
other pair, i.e. ($\nu_2$, $\nu_4$) and ($\nu_8$, $\nu_{10}$), have
{\it different} symmetry groups: the group $G=3a$ corresponds to $\nu_4$,
$\nu_8$ and $G =6a$ corresponds to $\nu_2$, $\nu_{10}$. 
Since both modes of each of the above mentioned pairs must appear
{\it simultaneously} (this can be seen from the structure of Eqs.~(\ref{eq54}a), 
(\ref{eq54}b), (\ref{eq54}d)
and (\ref{eq54}e), respectively), the bush $B[3a]$, in its pure form,
cannot be excited in connection with the loss of stability of the
bush $B[2a]$. 

Indeed, as was just stated, according to, for example,
equation (\ref{eq54}b), the excitation of the mode $\nu_4$ with $G=3a$
leads necessarily to excitation of the root mode $\nu_2$ ($G=6a$)
of the {\it larger} bush $B[6a]$.
Thus, equations (\ref{eq54}b)
\begin{eqnarray}\label{eq58}
\ddot\nu_2+\omega_2^2\nu_2&=&-2(\alpha A)\nu_4\cos(2\tau)\,,\\
\ddot\nu_4+\omega_4^2\nu_4&=&-2(\alpha A)\nu_2\cos(2\tau)\,\nonumber
\end{eqnarray}
describe the transition from the original bush $B[2a]$ to the bush 
$B[6a]$ [root modes $\nu_2(t)$, $\nu_{10}(t)$], when $(\alpha A)$ 
reaches the appropriate critical value $(\alpha A)_c$. Direct 
numerical computations for Eqs.~(\ref{eq58}) lead to 
the following result:
\begin{equation}\label{eq59}
(\alpha A)_c=1.049\,.
\end{equation}
Comparing the critical values (\ref{eq56*}) and (\ref{eq59}) of 
$(\alpha A)$ providing the transition to the bushes $B[4a]$ and $B[6a]$ 
respectively, we see that these bushes  are excited 
by the mode $\nu_6(t)$ at the same threshold (up to the numerical
accuracy).

Similar to the above analysis, we can consider the stability of
the bush $B[2a]$ with respect to interactions with the root modes 
of the bush $B[12a]$. There are four such modes: $\nu_1, \nu_5$ and 
$\nu_7, \nu_{11}$ conjugate to them. Nevertheless, it is sufficient 
to take into account only two equations (\ref{eq54}a) for $\nu_1$ 
and $\nu_5$ because Eqs.~(\ref{eq54}d) for $\nu_7, \nu_{11}$ can be 
transformed to the form (\ref{eq54}a). The critical value 
$(\alpha A)_c$ which corresponds to the transition $B[2a]\to B[12a]$ 
also coincides with the above values (\ref{eq56*}) and (\ref{eq59}) for
transitions  $B[2a]\to B[4a]$ and $B[2a]\to B[6a]$:
\begin{equation}\label{eq60}
(\alpha A)_c=1.049\,.
\end{equation}
We will return to this surprising coincidence in the next
 section of the present paper.

Comparing all three variants [see Eqs.~(\ref{eq56*}), (\ref{eq59}), 
(\ref{eq60})] of the loss of the stability of the original bush $B[2a]$,
which brought about its extension  up to the bushes $B[4a]$, $B[6a]$ 
and $B[12a]$, respectively, we find  that as the 
amplitude A of the mode $\nu_6(t)$ is increased,  the transition $B[2a]\to 
B[12a]$ 
must occur because the bush $B[12a]$
embraces both bushes $B[4a]$ and $B[6a]$.
This result, obtained from approximate Eqs. (\ref{eq54}), was 
confirmed by direct numerical computation of the exact dynamical 
equations for the FPU-$\alpha$ model. 
We wrote two variants of the appropriate computer programs:
the program P1 solves Eqs.~(\ref{eq28}) 
in ordinary space and then decomposes the vector ${\bf X}(t)$, 
defined by Eq.~(\ref{eq2}), into the set of modes, while the 
program P2 solves Eqs.~(\ref{eq32}) for the FPU-$\alpha$ chain in
modal space. Both programs give the same estimation for the value 
of $(\alpha A)_c$:
$$1.04<(\alpha A)_c<1.05\;,$$
and confirm the above conclusion that the bush $B[2a]$ 
really transmutes into the bush $B[12a]$ as a result of the loss 
of its stability.

Below we give the critical values of $(\alpha A)$ for the loss of 
stability of all other bushes for the FPU-$\alpha$ chain with $N=12$:
\begin{eqnarray}\label{eq70}
B[4a]&\to& B[12a]:(\alpha A)_c\approx 1.05\,,\\
B[3a]&\to& B[6a]:(\alpha A)_c\approx 0.60\,,\label{eq71}\\
B[6a]&\to& B[12a]:(\alpha A)_c\approx 1.29\,.\label{eq72}
\end{eqnarray}

Let us pay attention to the loss of the stability of the bush 
$B[3a]$. As follows from Eq.~(\ref{eq71}), this bush
transmutes at $(\alpha A)_c=0.60$ into the embracing bush $B[6a]$.  
This transformation is accompanied by the lowering of symmetry 
$(G[3a]\supset G[6a])$ twice. The new bush $B[6a]$, arising as a result 
of this process, exists as an individual stable object for the following interval
of its own root mode
\begin{equation}\label{eq73}
0.60<\alpha A<1.29\,.
\end{equation}
In turn, the bush $B[6a]$ loses stability for 
$(\alpha A)>(\alpha A)_c=1.29$ and transmutes into the largest 
bush B[12a] containing all modes $\nu_j\; (j=1,2,\ldots,11)$, also with 
lowering of symmetry twice $(G[6a]\supset G[12a])$.\footnote{Transition from one bush $B_1$
to another bush $B_2$ with higher dimensionality (with more degrees of freedom) as
a result of the loss of stability of $B_1$ is always associated with spontaneous breaking
of symmetry of the original dynamic regime described by the bush $B_1$. This is 
one of the central conclusions of the bush theory since every bush possesses its own symmetry 
group \cite{l7, l8}. In connection with this, let us note that some aspects of the loss 
of permutational symmetry of the vibrating chains are considered in~\cite{l27}.}

Thus, we have considered the stability of all possible bushes for the
FPU-$\alpha$ chain with $N=12$ particles.\footnote{
In this paper, we consider stability of the bushes associated with 
the translational parent group $G_0 = T$ only.
} All other cases with 
sufficiently small $N$ can be studied by similar methods. But some 
new phenomena appear when $N$ becomes very large, and we 
consider them in the next section.

\subsection{Stability of bushes of modes for large N}

Let us consider the stability of the bush $B[2a]$ for an arbitrary even
value of  $N$  (its stability for the case $N=12$ was discussed in 
Section~3.1). This bush consists of only one mode with $j=N/2$ 
and the appropriate dynamical equation 
\begin{equation}\label{eq74}
\ddot\nu_{N/2}+\omega_{N/2}^2\nu_{N/2}=0\;
\end{equation}
can be obtained from 
Eqs.~(\ref{eq32}) assuming that all modes different from $\nu_{N/2}(t)$
are equal to zero.
This is equation of the harmonic oscillator with $\omega_{N/2}=2$
[see Eq.~(\ref{eq30})], and we can write its solution in the form:
\begin{equation}\label{eq75}
\nu_{N/2}(\tau)=Acos(2\tau)\;.
\end{equation}
Linearizing the system (\ref{eq32}) near the exact solution 
(\ref{eq75}) leads to the following approximate equations:
\begin{eqnarray}\label{eq76}
&&\ddot\nu_j+\omega_j^2\nu_j=-\frac{8\alpha}{\sqrt {N}}
\sin\left(\frac{2\pi j}{N}\right)\nu_{N/2}\;\nu_{N/2-j}\;,\\
&&\omega_j^2=4\sin^2\left(\frac{\pi j}{N}\right),\;\;
j=1,2,\ldots,\frac{N}{2}-1,\frac{N}{2}+1,\ldots,N-1\;.\nonumber
\end{eqnarray}
Note, that right-hand sides of these equations contain only the term 
$\nu_{N/2}\;\nu_{N/2-j}$ (the coefficients before similar terms
$\nu_{N/2}\;\nu_{N/2+j}$ turns out to be zero). It is easy to see that
the system (\ref{eq76}) splits into a number of subsystems containing
one or two equations only. Indeed, the mode $\nu_j$ in (\ref{eq76})
is connected with the mode $\nu_{\tilde j}$ where
$\tilde j=\frac{N}{2}-j$, and vice versa, the mode $\nu_{\tilde j}$ 
is connected with $\nu_{\frac{N}{2}-\tilde j}\equiv\nu_j$.
Therefore, we have for $j=1,2,\ldots,\frac{N}{2}-1$ in Eq.(\ref{eq76})
the following pairs of equations which are 
{\it independent} from all other equations:
\begin{eqnarray}\label{eq77}
\ddot\nu_j+4\sin^2\left(\frac{\pi j}{N}\right) \nu_j =
-\gamma\sin\left(\frac{2\pi j}{N}\right)\nu_{\frac{N}{2}-j}\cos(2\tau)\;,\\
\ddot\nu_{\frac{N}{2}-j}+4\cos^2\left(\frac{\pi j}{N}\right)
\nu_{\frac{N}{2}-j}=-\gamma\sin\left(\frac{2\pi j}{N}\right)
\nu_j\cos(2\tau)\;,\nonumber
\end{eqnarray}
where
\begin{equation}\label{eq78}
\gamma=\frac{8\alpha A}{\sqrt{N}},\;\;j=1,2,\ldots,(\frac{N}{4}-1)\;.
\end{equation}
Here we made use of the explicit form of $\nu_{N/2}(t)$ from  
Eq.~(\ref{eq75}). For $j=\frac{N}{2}+1,\ldots,N-1$
we obtain from (\ref{eq76}) the equations similar to (\ref{eq77}),
 but with the changing 
of the sign of the constant $\gamma$. As was discussed in
the previous section, this case can be reduced to Eqs.~(\ref{eq77})
by replacing $\tau\to\tau+\frac{\pi}{2}$.
(Note, that for $j > \frac{N}{2}$, the index $\tilde j=\frac{N}{2}-j$
becomes negative and then we can use the cyclic condition $j+N\equiv j$
for obtaining positive values of $\tilde j$). 
When $j=\frac{N}{2}-j$, both modes $\nu_j$ and $\nu_{\frac{N}{2}-j}$
reduce to the same mode $\nu_{\frac{N}{4}}$ and (\ref{eq77}) reduces 
to the pair of identical equations of the form
\begin{equation}\label{eq79}
\ddot\nu_{\frac{N}{4}}+2\nu_{\frac{N}{4}}=-\gamma\,\nu_{\frac{N}{4}}
\cos(2\tau)\;.
\end{equation}
Obviously, this equation can be transformed to the standard 
form of the Mathieu equation (\ref{eq56}) with $y(\tau)=\nu_{\frac{N}{4}}(\tau), 
a=2, q=-\gamma/2$.

Thus, the system (\ref{eq76}) of $N-2$ differential equations 
splits into the subsystems (\ref{eq77}) or (\ref{eq79}) according 
to the diagram shown in Fig. 1, where indices
$j$ corresponding to pairs of equations (\ref{eq77}) are connected with 
the arches of different size.

It is convenient to rewrite Eqs.~(\ref{eq77}) 
 as follows:\footnote{
 As  was already noted, the sign before $\gamma $ 
in Eqs.~(\ref{eq77}) is not essential.}:
\begin{eqnarray}\label{eq80}
\ddot x+4\sin^2(k)x=\gamma\sin(2k)y\cos(2\tau)\;,\\
\ddot y+4\cos^2(k)y=\gamma\sin(2k)x\cos(2\tau)\;.\nonumber
\end{eqnarray}
where $k={\pi j}/N, x(\tau)=\nu_j(\tau), y(\tau)=\nu_{\frac{N}{2}-j}(\tau)$.
Thus, studying the loss of stability of the bush $B[2a]$, 
brought about by its interactions with the modes $\nu_{N/4}$
and $\nu_{3N/4}$ is reduced to analyzing the Mathieu equation (\ref{eq79}),
and the loss of its stability with respect to all other modes reduces
to analyzing Eqs.~(\ref{eq80}).

We begin to study the stability of the bush $B[2a]$ by
considering its interaction with the mode $\nu_{N/4}$ 
[let us remember that $B[2a]$ consists of only one mode $\nu_{N/2}$
described by Eq.~(\ref{eq75})]. It is clear that $B[2a]$ 
must be a stable object for very small values of $\alpha A$
and we are interested in the lower boundary $A_c$ 
of the loss of its stability, because of the parametric excitation of 
the mode $\nu_{N/4}$, when the amplitude $A$  of the mode $\nu_{N/2}$
increases from zero. This problem can be solved easily with the 
aid of the Mathieu equation (\ref{eq79}). Using its well known 
stability chart and analytical formulas for boundaries of unstable 
regions in the $(a,q)$ plane \cite{l19}, and independently by direct numerical 
investigation of Eq.~(\ref{eq79}), we found the following threshold value $\gamma_c$ 
for the coefficient $\gamma$ entering into this equation:
\begin{equation}\label{eq81}
\gamma_c=2.42332\;.
\end{equation}
Then from Eq.~(\ref{eq78}) we obtain:
\begin{equation}\label{eq82}
\alpha A_c = 0.30292\sqrt{N}\;.
\end{equation}
Thus, the bush $B[2a]$ is stable with respect to the interaction 
with the mode $\nu_{N/4}$ for $\mid \alpha A\mid <\alpha A_c$ 
given by Eq.~(\ref{eq82}). It is obvious that the same value of 
$\alpha A_c$ corresponds to the loss of stability of 
$B[2a]$ because of parametric excitation of the mode $\nu_{3N/4}$.

Now we must take into account interactions of the bush $B[2a]$ 
with all other modes $\nu_j(t)$. As was already stated, studying the
stability of $B[2a]$ leads, in this case, to the equations (\ref{eq80}). 
They are linear differential equations with periodic coefficients 
and, therefore, the Floquet theory can be applied to them. 
Unfortunately, we did not obtain any exact solutions of 
Eqs.~(\ref{eq80}) and can now give only results of the appropriate 
numerical analysis. Our computation of the multiplicators for 
the system (\ref{eq80}) as eigenvalues of the monodromic matrix 
reveals a surprising fact! Indeed, it seems that the critical value 
$\gamma_c$ of the constant $\gamma$ from (\ref{eq77}) 
[see also Eqs.~(\ref{eq80})], corresponding to the loss of stability
of the bush $B[2a]$, must depend on the mode number $j$, since the coefficients 
of these equations depend explicitly on $j/N$. However, this is 
not true. We found that $\gamma_c$ does not depend on $j/N$, 
at least up to $10^{-5}$, and coincides with that 
of the Mathieu equation (\ref{eq79}): 
$$\gamma_c=2.42332.$$

This nontrivial fact means that the original bush $B[2a]$ loses 
its stability with respect to all modes $\nu_j$ $(j \not= N/2)$ 
{\it simultaneously}, i.e., for the same value $A_c$ of the amplitude of 
the mode $\nu_{N/2}(t)$. In other words, all modes of an $N$-particle
FPU-$\alpha$ chain are excited parametrically because of 
interaction with $B[2a]$ when $\alpha A$ reaches its critical 
value $\alpha A_c~=~0.30292\sqrt{N}$ and, therefore, 
$B[2a]$ transforms at once into the bush $B[Na]$ of trivial 
symmetry.\footnote{
We already met this phenomenon [see Eqs.(\ref{eq56*}),(\ref{eq59}) and (\ref{eq60})]
for the special case $N=12$. The value $\alpha A_c = 1.049$ from the above mentioned
equations is obtained from our present formula $\alpha A_c = 0.30292\sqrt{N}$
for $N=12$.}

The above results obtained by studying Eqs.~(\ref{eq77},\ref{eq79}) 
were verified with the aid of the program 
P1 for solving the dynamical equations (\ref{eq28}) for the atomic 
displacements $x_i(t)$, $i=1,2,\ldots,N$. Indeed, for $N=50,100,200$, 
we obtained in such a way the relation $\alpha A_c = 0.303\sqrt{N}$ which is 
in the excellent agreement with Eq.~(\ref{eq82}). Moreover, it was also 
found that {\it all} modes $\nu_j(t),j=1,2,\ldots,N-1$ are present in the 
decomposition (\ref{eq20}) of the configurational vector
${\bf X}(t)$ from 
Eq.~(\ref{eq1}). This fact confirms the above discussed conclusion about
the appearance of the bush $B[Na]$ from the original bush $B[2a]$ 
when the amplitude $A$ exceeds the critical value $A_c$ from 
Eq.~(\ref{eq82}).

 Note that using the program P1 for studying the 
stability of the bush $B[2a]$, we try N only up to
1024, but analyzing this problem with the aid of Eqs.~(\ref{eq77})
 we try very great values of $N$, such as $N=10^4,10^5$ allowing $j/N$ to approach
 sufficiently close to the ``dangerous" zero value.

Now let us consider the configuration vector ${\bf X}(t)$ from
Eq.~(\ref{eq20}) which determines displacements of all $N$ particles 
from their equilibrium states expressed in terms of the 
modes $\nu_j(t)$. The bush $B[2a]$ consists of only one mode 
$\nu_{N/2}$ and, therefore,
\begin{equation}\label{eq83}
{\bf X}(t)\Big |_{B[2a]}=\nu _{N/2}(t)\,\mbox{\boldmath$\psi $}_{N/2}\;.
\end{equation}
The components $(\mbox{\boldmath$\psi $}_{N/2} )_n$
of the N-dimensional vector $\mbox{\boldmath$\psi $}_{N/2}$ 
can be obtained from Eq.~(\ref{eq19}):
\begin{equation}\label{eq84}
(\mbox{\boldmath$\psi $}_{N/2} )_n= \frac{1} {\sqrt{N}} \cos (\pi n),
\;\;n=1,2,\ldots,N.
\end{equation}
Then we have
\begin{equation}\label{eq85}
\mbox{\boldmath$\psi $}_{N/2} = \frac{1} {\sqrt{N}}\,{\bf c},\;
\mbox{where} \;{\bf c}=(-1,1,-1,1,\ldots,-1,1)
\end{equation}
and
\begin{equation}\label{eq86}
{\bf X}(t)\Big |_{B[2a]}=\frac{A} {\sqrt{N}}\cos (2t)\;{\bf c}.
\end{equation}
Here, for simplicity, we choose such an initial condition for the excitation 
of the bush $B[2a]$ that $\delta$ from Eq.~(\ref{eq53}) is equal to zero.
It is clear from (\ref{eq85}), (\ref{eq86}) 
that the displacement pattern of the FPU chain, corresponding 
to arbitrary instant $t$, possesses the 
translational symmetry $2a$ (see Fig.2), since displacements of 
every two particles $2a$ apart from each other are 
identical.

The dynamical regime (\ref{eq86}) loses its stability when the 
amplitude $A$ becomes larger then $A_c$ determined by 
Eq.~(\ref{eq82}):
\begin{equation}\label{eq87}
A_c\approx \frac{0.303}{\alpha }\;\sqrt N.
\end{equation}

Substituting this value into Eq.~(\ref{eq86}) we obtain 
the {\it critical} form ${\bf X}(t)$ for the considered bush $B[2a]$:
\begin{equation}\label{eq88}
{\bf X}_c(t)\Big |_{B[2a]}=\frac{0.303}{\alpha }\;
{\bf c}\;\cos (2t).
\end{equation}

Note that ${\bf X}_c(t)$ from (\ref {eq88}) 
does not depend on the number of particles $(N)$ in
the FPU chain! It follows from (\ref {eq88}) that 
amplitudes of all atomic displacements for critical
 ${\bf X}_c(t)$ must be larger for lesser values of
 $\alpha$ and, formally, they tend to 
infinity when $\alpha \to 0$. Such a behavior reflects the fact 
that there are no interactions between normal modes in the limit 
$\alpha=0$ and, therefore, the normal mode $\nu_{N/2}(t)$ 
becomes stable in this case for any arbitrary amplitude 
$A$. To avoid a possible misunderstanding, let us note that for a
FPU-$\alpha$ system, considered as an abstract {\it exact model}, 
any distance $a$ between particles in equilibrium is permitted 
and, as a result, atomic displacements can be large.
From this point of view, the presence $\alpha$ in the 
denominator in Eq.~(\ref {eq88}) must not be striking. 
However, when we consider the FPU-$\alpha$ model as 
an approximation to the real physical system, only sufficiently
small atomic displacements are assumed because we
must neglect, in such a case, the higher terms in the Taylor 
series (\ref {eq0}) for $F(\Delta x)$. 
We must also take into account that the value of the amplitude 
$A$ of the {\it oscillatory} regime is restricted by the presence of a 
potential barrier which appears because the nonlinear term with 
$\alpha$ in the FPU-$\alpha$ Hamiltonian (\ref {eq27}) can be negative 
with absolute value larger than the quadratic term (the height of this 
barrier and its distance from zero are proportional to $1/\alpha^2$ and 
$1/\alpha$, respectively).

There was a discussion on the loss of stability of the zone-boundary mode (ZBM)
in \cite{l21, l22, l20}.\footnote{
We are very grateful for the referee of our paper for these references which we
did not know previously.}
In our terms, this problem is identical to the problem of the loss of stability
of the bush $B[2a]$, since this bush consists of only one above mentioned mode.\footnote{
In some papers, the zone-boundary mode is called by the term ``$\pi$-mode".}
In the present paper, we do not want to give detailed comparisons of the
results by different authors and restrict ourselves
by the following short comments.

In the paper \cite{l20} by Sandusky and Page, the
interrelation between the loss of stability of the zone-boundary 
mode (ZBM) and existence of
intrinsic localized modes (ILM's)\footnote{
The term ``breathers" is usually used for ILM's at the 
present time.} 
is discussed for different nonlinear monatomic chains.
It was found that there can exist two different
types of the loss of stability of the ZBM in the ($k_2, k_3, k_4$)
chains.\footnote{Here, $k_2, k_3, k_4$ are coefficients in front of the 
harmonic term ($k_2$) and in front of the anharmonic terms of the third 
($k_3$) and fourth ($k_4$) orders.}
One of these types is connected with the appearance of
the ILM's, while the other, the so called ``period-doubling" type\footnote{
This term means that the new mode which appears 
because of the loss of stability of the ZBM
possesses a time period two times larger than that of ZBM.},
is not related to ILM's.

Only the ``time-doubling" type of the loss of stability
of ZBM occurs when the coefficient $k_3$ is
sufficiently large, in particular, when $k_3 \not = 0$, $k_4 = 0$
(this case corresponds to the FPU-$\alpha $ model).
Namely this phenomenon is discussed in the present
paper for the FPU-$\alpha $ chain, and we can compare
the values of the amplitude threshold for the
loss of stability of the ZBM (the bush $B[2a]$, 
in our terms) obtained by us and by Sandusky
and Page. The following estimate was found in [19]:
\begin{equation}\label{eq245}
0.302 < \mid \Lambda _3\mid  < 0.303,
\end{equation} 
where $\Lambda _3= k_3 A /k_2$ with $A$ being the non-normalized
amplitude of the ZBM.

Taking into account Eq.(\ref{eq27}), we can
express $\Lambda _3$ in terms of our paper:
\begin{equation}\label{eq246}
\Lambda _3 = \alpha  A / \sqrt{N},
\end{equation} 
where $A$ is the mode amplitude with respect to {\it normalized} basis vectors (\ref{eq19}).
Then, with the aid of the Eq.~(\ref{eq82}) 
$$
\alpha  A_c = 0.30292 \sqrt{N}
$$
we obtain
\begin{equation}\label{eq247}
\Lambda _3= 0.30292
\end{equation} 
This value of the threshold of the loss of stability of the ZBM is in the excellent
agreement with estimation (\ref{eq245}) obtained in the paper [19] for the case $N=100$.
Our result (\ref{eq247}) is valid for arbitrary value of $N$ and, moreover, as was already
discussed in the text of the present paper, we found that all modes appear 
{\it simultaneously}\footnote{We suspect that this phenomenon leads, as a final result, to
equipartition of the energy between different modes of the FPU-$\alpha$ chain, but we did
not study this fact.}
when $\Lambda _3$ exceeds the remarkable threshold (\ref{eq247}). This threshold was obtained
by another method, as compared to [19], and with the higher precision.

Thus, we confirmed the validity of the results on the loss of  the ZBM stability for the FPU-$\alpha $
chain by Sandusky and Page from the paper [19] in contrast to those 
obtained in [20,21] (the discussion on the causes of the incorrectness of the appropriate
results reported in [20,21] can be found in [19]).

In many papers, the energy density $\epsilon = E(0)/N$ of the initial excitation,
i.e. the energy $E(0)$ per one particle, is considered as a relevant control parameter
characterizing the loss of stability of different modes in nonlinear chains. Substituting
$X_c(0)$ from (\ref{eq88}) into the Hamiltonian (\ref{eq27}) leads to vanishing the sum
of anharmonic (cubic) terms, and we obtain the following expression for the threshold
$\epsilon_c$ of the loss of stability of the bush $B[2a]$:
$$
\epsilon = 0.18325/{\alpha^2}.
$$
Let us emphasize that this value does not depend on the number $N$ of the particles in
the FPU-$\alpha$ chain.

We want to note that unlike one-dimensional bushes, the energy density $\epsilon$
is not a relevant parameter for characterizing bush stability in the general case. Indeed, 
it can be shown that if a given bush contains several modes, the threshold of the loss
of stability depends not only on $\epsilon$ but on the {\it distribution} of the energy among
its modes at the initial instant. Because of this reason, we prefer to fix this initial energy
distribution in a certain way, namely, we assume that the energy is completely localized
only in the {\it root} mode of the bush and then we look for the threshold value $A_c$ of this mode.

In conclusion, let us note that the rigorous stability analysis of all bushes of 
vibrational modes cannot be fulfilled on the basis of the Floquet theory\footnote{
 Note that the threshold of the loss of stability of
the ZBM, i.e. of the bush $B[2a]$, was obtained in \cite{l20} and in the
present paper with the aid of the Floquet theory.},
because, in general, dynamical regimes corresponding to them are not {\it periodical}.
Indeed, a given bush can contain modes with {\it incommensurable} frequencies
(for example, the bush $B[4a]$ contains the modes with the harmonic frequencies
$\omega_{3,9} = \sqrt{2}$ and $\omega_6 = 2$).

Now let us compare our results on stability of the bush $B[2a]$ 
for the FPU-$\alpha$ model with those by Poggi and Ruffo obtained 
in \cite{l3} for the FPU-$\beta$ model [remember that $B[2a]$ is the 
simplest bush consisting of only one mode $\nu_{N/2}(t)$]. The 
dynamical equation describing the evolution of the mode $\nu_{N/2}(t)$
is the equation of the harmonic oscillator for the FPU-$\alpha$ chain
and the Duffing equation for the FPU-$\beta$ chain, respectively. The 
problem of the loss of stability of the bush $B[2a]$ according to 
the parametric excitation of the other modes reduces in the 
linear approximation to studying the differential equations 
(\ref{eq77},\ref{eq79}) for the FPU-$\alpha$ case and to 
studying the Lam\'e 
equation for the FPU-$\beta$ case. In both cases, coefficients 
of the above mentioned equations depend on $j/N$ only, where 
$j$ is a number of the excited mode and $N$ is the total 
number of particles in the FPU chain.

Let us consider the main difference in stability properties of the 
bush $B[2a]$ for the two discussed mechanical systems. There are 
different threshold values of the energy localized in the mode 
$\nu_{N/2}$ for exciting the different modes $\nu_j$ for the 
FPU-$\beta$ chain and, moreover, the modes for which $\sin^2(\frac{\pi 
j}{N})<\frac{1}{3}$ cannot be excited parametrically at all \cite{l3}. In 
contrast to this, we found that in the FPU-$\alpha$ model {\it all modes}
with $j \not = \frac{N}{2}$ are excited by the mode 
$\nu_{N/2}$ {\it simultaneously}, i.e., the same threshold value 
$\alpha A_c$ corresponds to them. When $\mid \alpha A\mid >\alpha A_c$
 all modes $\nu_j (j \not= \frac{N}{2})$ begin to increase at once,
and, as a result, the original bush $B[2a]$ loses its stability.

The studying of stability of the bushes other than $B[2a]$ for 
arbitrary $N$ (which are multidimensional) is the more 
complicated problem, and we will not discuss it in the present 
paper. Let us point out only one interesting result. 
As was discussed in Sec. 2, every bush 
possesses a certain symmetry group $G$ and contains a number 
of modes whose symmetry is higher or equal to $G$. These modes 
$\nu_j$ are depicted as sticks in Fig.3a for the case of the 
bush $B[8a]$ and we will refer to them, using the spectroscopic 
terminology, as ``lines" (note that some modes of the bush
are too small to be shown on this picture). 
The height of each such stick (line) is 
determined by the value of $\nu_j(t)$ at the instant $t$. When 
$N$ becomes sufficiently large (using the program P1, we 
considered the cases $N=128, 256, 512, 1024$) a number of 
satellites, forbidden by the above mentioned symmetry 
restriction, appear near every line of a given bush (see Fig.~3b) for 
the amplitude of the root mode exceeding a certain critical value. 
This phenomenon may be described in the following manner.

Let $j_0$ correspond to one of the modes $\nu_{j_0}$ belonging 
to a given bush. Then several ``forbidden" modes $\nu_j$ with 
$j=j_0\pm 1, j_0\pm 2, j_0\pm 3,\ldots$ arise above
 a certain threshold for 
$\alpha A_c$ of a root mode. It must be stressed that the 
appearance of such ``satellites" of different orders is not 
connected with any symmetry related principles. The excitation 
of the satellites brought about by the approach of frequencies of 
the appropriate modes to each other when $N \to \infty$ and this phenomenon,
in turn, results in resonance interactions between modes. 
The time evolution of the satellites can 
be very nontrivial, in particular for large $N$, amplitudes of satellites 
of higher orders can be greater than those of lesser orders. 
In Fig 3.b, we show schematically the satellite structure of the bush 
$B[8a]$ for N=128 at $t\approx 2500 T_L$, where $T_L$ is the
period of oscillations of the root mode. Note, that ``first" modes, i.e. 
modes $\nu_j$ with $j$  near the beginning and near the end of 
the interval of permissible values of mode numbers, are
also present on the Fig.3.b. The appearance of the satellite structure
corresponds to the beginning of the process of the loss of stability 
of the given bush, but this structure can be observed for a sufficiently
long time as a metastable state.

In connection with the above mentioned metastable states, let us note that 
there are a number of papers devoted to studying the dynamics of nonlinear
chains {\it above} the threshold of the loss of stability of the
zone-boundary mode, i.e. the bush $B[2a]$, in our terms (see, for example,
\cite{l20} and \cite{l28} where some types of localized states were revealed
in the certain cases). In particular, the very interesting phenomenon of the emergence of
``chaotic breathers" as a kind of metastable dynamical regime was reported in
\cite{l28}. We do not study such phenomena in the present paper.

\section{Conclusion}

In the present paper we demonstrate how to obtain the 
invariant submanifolds in the modal space of an $N$-particle chain 
with periodic boundary conditions by regular group-theoretical 
methods based on the concept of bushes of normal modes 
developed by us in [7-11] for {\it arbitrary} mechanical systems with 
discrete symmetry. Bushes of modes appear under 
{\it certain initial conditions} and they can be considered as geometrical 
objects with their own symmetry and as dynamical objects, as well.
For the last case, differential equations 
corresponding to the dynamical 
systems whose dimensionality is frequently essentially less than $N$ can 
be obtained. The energy localized in the very definite set of 
modes, corresponding to a given bush, remains trapped. The 
object of such a type were revealed recently for the FPU-$\beta$ 
chain by Poggi and Ruffo \cite{l3}. We show that these objects do 
correspond to our concept of bushes of normal modes and that 
the group-theoretical methods developed by us earlier allow 
one to find easily all these objects for arbitrary chains 
(not for the FPU models only). 
We also compare the classification of the bushes of vibrational modes according
with the different choice of the parent symmetry group $G_0$
(it can be the cyclic group of pure translations or the dihedral group).

For a concrete chain---FPU-$\alpha$ model---we derive 
the integro-differential equation which 
describes the dynamics of this chain in the modal space in the 
limit $N \to \infty$. In contrast to the KdV equation obtained by 
Zabusky and Kruskal for the FPU-$\alpha$ model in the continuum limit 
in the ordinary space, the above mentioned integro-differential 
equation is correct for all wave lengths, not only for long waves. 

We study the stability of all possible bushes of normal modes 
for the case $N=12$. For the bush $B[2a]$ to which the pattern 
of atomic displacements with translational symmetry twice 
that of the equilibrium state corresponds, we found the 
boundary of the loss of stability in the limit $N \to \infty$. It was 
found that the bush $B[2a]$ loses its stability with respect to all 
other modes {\it simultaneously} when the amplitude of atomic 
displacements exceeds the value $(0.303/\alpha)$. Let us stress 
that this result corresponds only to the FPU-$\alpha$ model. Indeed, 
Poggi and Ruffo found essentially different results for the 
FPU-$\beta$ chain \cite{l3} (we give the appropriate comparison in
Sec. 3.2 of the present paper). We also briefly discussed the 
appearance of the metastable satellite structure of symmetry forbidden lines 
near the permitted lines of the bush $B[ma]$ where $m \not = 2$ 
for sufficiently large $N$.\\

{\Large {\bf Acknowledgments}}\rule{0pt}{30pt}

The \rule{0pt}{30pt}authors thank Prof. V.P. Sakhnenko for useful discussion and
 Prof. H.T. Stokes for many remarks and corrections in the text
of the present paper. We appreciate this help very much. G.M. Chechin
thanks Profs. D.M. Hatch and H.T. Stokes for their hospitality
during his visit to Brigham Young University in summer 1999. This 
visit stimulated the preparation of this paper for publication.
We thank our referees for very constructive criticism which helped us to
improve the paper essentially.

\clearpage
\begin{figure}
\begin{picture}(420,160)
\put(20,20){
\begin{picture}(380,100)
\put(0,20){\line(1,0){370}}
\put(185,20){\line(0,1){50}}
\put(97.5,20){\line(0,1){50}}
\put(272.5,20){\line(0,1){50}}
\multiput(10,20)(12.5,0){29}{\circle*{3}}
\put(185,20){\circle{8}}

\put(-10,4){$j:$}
\put(93,4){$\frac{N}{4}$}
\put(180.5,4){$\frac{N}{2}$}
\put(266,4){$\frac{3N}{4}$}
\put(8,4){$\scriptstyle 0$}
\put(20.5,4){$\scriptstyle 1$}
\put(33,4){$\scriptstyle 2$}
\put(356,4){$N$}

\put(77.5,90){\it Mathieu}
\put(77.5,80){\it equation}
\put(252.5,90){\it Mathieu}
\put(252.5,80){\it equation}
\put(155,100){\it The equation}
\put(157,90){\it of harmonic}
\put(165,80){\it oscillator}

\put(97.5,20){\oval(25,10)[t]}
\put(97.5,20){\oval(50,20)[t]}
\put(97.5,20){\oval(75,30)[t]}
\put(97.5,20){\oval(100,40)[t]}
\put(97.5,20){\oval(125,50)[t]}
\put(97.5,20){\oval(150,60)[t]}

\put(272.5,20){\oval(25,10)[t]}
\put(272.5,20){\oval(50,20)[t]}
\put(272.5,20){\oval(75,30)[t]}
\put(272.5,20){\oval(100,40)[t]}
\put(272.5,20){\oval(125,50)[t]}
\put(272.5,20){\oval(150,60)[t]}

\end{picture}}

\put(0,5){\footnotesize Fig.1. Splitting of Eqs. (\ref{eq76}).}

\end{picture}
\end{figure}

\begin{figure}
\begin{picture}(420,150)
\put(100,20){

\begin{picture}(100,100)

\put(10,90){\line(1,0){170}}
\multiput(20,30)(60,0){3}{\vector(-1,0){10}}
\multiput(20,90)(30,0){6}{\circle*{5}}
\multiput(50,70)(30,0){2}{\line(0,1){20}}
\put(50,75){\vector(1,0){30}}
\put(80,75){\vector(-1,0){30}}
\put(63,76.5){$a$}

\put(10,30){\line(1,0){170}}
\multiput(20,30)(30,0){6}{\circle*{5}}
\multiput(20,30)(60,0){3}{\vector(-1,0){10}}
\multiput(20,30.3)(60,0){3}{\line(-1,0){8}}
\multiput(20,29.7)(60,0){3}{\line(-1,0){8}}
\multiput(50,30)(60,0){3}{\vector(1,0){10}}
\multiput(50,30.3)(60,0){3}{\line(1,0){8}}
\multiput(50,29.7)(60,0){3}{\line(1,0){8}}

\multiput(35,33)(60,0){2}{\line(0,-1){23}}
\put(64,15){\vector(1,0){31}}
\put(66,15){\vector(-1,0){31}}
\put(59,16.5){$2a$}

\end{picture}}
\put(0,5){\footnotesize Fig.2. The monatomic chain in the equilibrium state
and the pattern of atomic displacements 
for the bush $B[2a]$.}
\end{picture}
\end{figure}

\begin{figure}
\begin{picture}(420,250)
\put(0,147){(a)}
\put(0,47){(b)}
\put(20,30){

\begin{picture}(380,220)
\put(0,20){\line(1,0){380}}
\put(0,120){\line(1,0){380}}

\put(49,120){\line(0,1){80}}
\put(96,120){\line(0,1){30}}
\put(143,120){\line(0,1){0}}
\put(190,120){\line(0,1){0}}
\put(237,120){\line(0,1){10}}
\put(284,120){\line(0,1){31}}
\put(331,120){\line(0,1){0}}

\put(49,20){\line(0,1){80}}
\put(96,20){\line(0,1){38}}
\put(143,20){\line(0,1){8}}
\put(190,20){\line(0,1){3}}
\put(237,20){\line(0,1){12}}
\put(284,20){\line(0,1){32}}
\put(331,20){\line(0,1){55}}

\put(0,20){\begin {picture}(0,0)
\put(0,0){\line(0,1){0}}
\put(3,0){\line(0,1){31}}
\put(6,0){\line(0,1){13}}
\put(9,0){\line(0,1){2}}
\put(371,0){\line(0,1){1}}
\put(374,0){\line(0,1){15}}
\put(377,0){\line(0,1){30}}
\end{picture}}

\put(37,20){\begin {picture}(0,0)
\put(0,0){\line(0,1){0.5}}
\put(3,0){\line(0,1){2}}
\put(6,0){\line(0,1){8}}
\put(9,0){\line(0,1){25}}
\put(15,0){\line(0,1){21}}
\put(18,0){\line(0,1){7}}
\put(21,0){\line(0,1){1}}
\end{picture}}

\put(84,20){\begin {picture}(0,0)
\put(0,0){\line(0,1){0.6}}
\put(3,0){\line(0,1){1}}
\put(6,0){\line(0,1){4}}
\put(9,0){\line(0,1){13}}
\put(15,0){\line(0,1){11}}
\put(18,0){\line(0,1){2}}
\put(21,0){\line(0,1){1}}
\end{picture}}

\put(131,20){\begin {picture}(0,0)
\put(0,0){\line(0,1){0.1}}
\put(3,0){\line(0,1){0.5}}
\put(6,0){\line(0,1){0.8}}
\put(9,0){\line(0,1){3}}
\put(15,0){\line(0,1){2}}
\put(18,0){\line(0,1){0.4}}
\put(21,0){\line(0,1){0.1}}
\end{picture}}

\put(178,20){\begin {picture}(0,0)
\put(0,0){\line(0,1){0.1}}
\put(3,0){\line(0,1){0.3}}
\put(6,0){\line(0,1){0.6}}
\put(9,0){\line(0,1){1}}
\put(15,0){\line(0,1){0.5}}
\put(18,0){\line(0,1){0.3}}
\put(21,0){\line(0,1){0.1}}
\end{picture}}

\put(225,20){\begin {picture}(0,0)
\put(0,0){\line(0,1){0.1}}
\put(3,0){\line(0,1){0.6}}
\put(6,0){\line(0,1){1}}
\put(9,0){\line(0,1){3}}
\put(15,0){\line(0,1){4}}
\put(18,0){\line(0,1){0.6}}
\put(21,0){\line(0,1){0.2}}
\end{picture}}

\put(275,20){\begin {picture}(0,0)
\put(21,0){\line(0,1){0.6}}
\put(0,0){\line(0,1){1}}
\put(3,0){\line(0,1){3}}
\put(6,0){\line(0,1){10}}
\put(12,0){\line(0,1){11}}
\put(15,0){\line(0,1){4}}
\put(18,0){\line(0,1){1.5}}
\end{picture}}

\put(322,20){\begin {picture}(0,0)
\put(21,0){\line(0,1){0.5}}
\put(18,0){\line(0,1){1}}
\put(15,0){\line(0,1){8}}
\put(12,0){\line(0,1){23}}
\put(6,0){\line(0,1){19}}
\put(3,0){\line(0,1){7}}
\put(0,0){\line(0,1){1}}
\end{picture}}

\end{picture}}
\put(0,15){\footnotesize Fig.3. Satellites near the modes of the
bush $B[8a]$ (schematically). (a) Absence of the satellite structure}
\put (0,3){\footnotesize  for $N=32$. (b) Appearance of the
 satellite structure for $N=128$.}
\end{picture}
\end{figure}

\clearpage

\begin{table}
\begin{flushleft}{\bf\tablename} {$\;${\bf1.} Irreducible
 representations of the  cyclic group $G_0$ for $N=12$}
\end{flushleft}
\begin{tabular}{|c|c|c|c|c|c|c|c|c|c|c|c|c|c|c|}
\hline 
Irreps&$E$&$a$&$a^2$&$a^3$&$a^4$&$a^5$&$a^6$&$a^7$&
$a^8$&$a^9$&$a^{10}$&$a^{11}$&\parbox{4em}
{\raisebox{-12pt}{$\!$Symmetry$\;$} \raisebox{-1pt}{groups $G_k$}}&Modes\\ 
\hline
$\Gamma _0$&$1$&$1$&$1$&$1$&$1$&$1$&$1$&$1$&$1$
&$1$&$1$&$1$&$[a]$&$\nu _0$\\
$\Gamma _1$&$1$&$\gamma $&$\gamma^2$&$\gamma^3$&$\gamma^4$&
$\gamma^5$&$\gamma^6$&$\gamma^7$&$\gamma^8$&$\gamma^9$&
$\gamma^{10}$&$\gamma^{11}$&$[12a]$&$\nu _1$\\
$\Gamma_2$&$1$&$\gamma^2$&$\gamma^4$&$\gamma^6$&$\gamma^8$&
$\gamma^{10}$&$1$&$\gamma^2$&$\gamma^4$&$\gamma^6$&
$\gamma^8$&$\gamma^{10}$&$[6a]$&$\nu _2$\\
$\Gamma_3$&$1$&$\gamma^3$&$\gamma^6$&$\gamma^9$&$1$&
$\gamma^3$&$\gamma^6$&$\gamma^9$&$1$&$\gamma^3$&
$\gamma^6$&$\gamma^9$&$[4a]$&$\nu _3$\\
$\Gamma_4$&$1$&$\gamma^4$&$\gamma^8$&$1$&$\gamma^4$&
$\gamma^8$&$1$&$\gamma^4$&$\gamma^8$&$1$&
$\gamma^4$&$\gamma^8$&$[3a]$&$\nu _4$\\
$\Gamma_5$&$1$&$\gamma^5$&$\gamma^{10}$&$\gamma^3$&$\gamma^8$&
$\gamma $&$\gamma^6$&$\gamma^{11}$&$\gamma^4$&$\gamma^9$&
$\gamma^2$&$\gamma^7$&$[12a]$&$\nu _5$\\
\hline
$\Gamma_6$&$1$&$\gamma^6$&$1$&$\gamma^6$&$1$&
$\gamma^6$&$1$&$\gamma^6$&$1$&$\gamma^6$&
$1$&$\gamma^6$&$[2a]$&$\nu _6$\\
$\Gamma _7$&$1$&$\gamma^7$&$\gamma^2$&$\gamma^9$&$\gamma^4$&
$\gamma^{11}$&$\gamma^6$&$\gamma^1$&$\gamma^8$&$\gamma^3$&
$\gamma^{10}$&$\gamma^5$&$[12a]$&$\nu _7$\\
$\Gamma_8$&$1$&$\gamma^8$&$\gamma^4$&$1$&$\gamma^8$&
$\gamma^4$&$1$&$\gamma^8$&$\gamma^4$&$1$&
$\gamma^8$&$\gamma^4$&$[3a]$&$\nu _8$\\
$\Gamma _9$&$1$&$\gamma^9$&$\gamma^6$&$\gamma^3$&$1$&
$\gamma^9$&$\gamma^6$&$\gamma^3$&$1$&$\gamma^9$&
$\gamma^6$&$\gamma^3$&$[4a]$&$\nu _9$\\
$\Gamma _{10}$&$1$&$\gamma^{10}$&$\gamma^8$&$\gamma^6$&$\gamma^4$&
$\gamma^2$&$1$&$\gamma^{10}$&$\gamma^8$&$\gamma^6$&
$\gamma^4$&$\gamma^2$&$[6a]$&$\nu _{10}$\\
$\Gamma _{11}$&$1$&$\gamma^{11}$&$\gamma^{10}$&$\gamma^9$&$\gamma^8$&
$\gamma^7$&$\gamma^6$&$\gamma^5$&$\gamma^4$&$\gamma^3$&
$\gamma^2$&$\gamma^1$&$[12a]$&$\nu _{11}$\\
\hline 
\multicolumn{15}{l}{\footnotesize Here $\gamma =e^{2\pi i/12}$ and, therefore,
$\gamma ^3=i$, $\gamma ^6=-1$, $\gamma ^9=-i$, $\gamma ^{12}=1$}\\
\end{tabular}
\end{table}

\begin{table}
\begin{flushleft}{\bf\tablename} {$\;${\bf2.} Bush classification in accordance with the
parent groups $G_0=T$ and $\tilde G_0 = D$}
\end{flushleft}
\begin{tabular}{|c|c|c|}
\hline 
Dim&$G_0=T$&$\tilde G_0 = D$ (new bush)\\
\hline
1&$[\Gamma _6]$&$[\Gamma _4 - \Gamma _8]$; $[\Gamma _3 - \Gamma _9]$\\
\hline
2&$[\Gamma _4, \Gamma _8]$&$[\Gamma _6, \Gamma _3 - \Gamma _9]$; $[\Gamma _2 - \Gamma _{10}, \Gamma_4 - \Gamma_8 ]$\\
\hline
3&$[\Gamma _6, \Gamma _3, \Gamma _9]$&$[\Gamma _6, \Gamma _2 - \Gamma _{10}, \Gamma _4 - \Gamma _8]$\\
\hline
5&$[\Gamma _6, \Gamma _2, \Gamma _{10}, \Gamma _4, \Gamma _8]$&
$[\Gamma _1 - \Gamma _{11}, \Gamma _2 - \Gamma _{10}, \Gamma _3 - \Gamma _9, \Gamma _4 - \Gamma _8, \Gamma _5 - \Gamma _7]$\\
\hline
6& - &$[\Gamma _6, \Gamma _1 - \Gamma _{11}, \Gamma _2 - \Gamma _{10}, \Gamma _3 - \Gamma _9, \Gamma _4 - \Gamma _8, \Gamma _5 - \Gamma _7]$\\
\hline
11&$[\Gamma _1, \Gamma _2, \ldots, \Gamma _{11}]$
& - \\
\hline
\end{tabular}
\end{table}

\end{document}